\documentclass[twocolumn,tighten,times]{aastex6}  
\usepackage{color}
\usepackage{graphicx}
\usepackage{mathrsfs}
\usepackage{ulem}
\usepackage{empheq}
\usepackage{multirow}
\usepackage{array}

\DeclareMathAlphabet{\mathbi}{OT1}{ptm}{bx}{it}
\SetMathAlphabet\mathbi{bold}{OT1}{ptm}{bx}{it}

\def\bhm{M_{\bullet}}

\def\ergs{\rm erg~s^{-1}}

\def\mathdotM{\dot{\mathscr{M}}}
\def\sunm{M_\odot}
\def\mathdotM{\dot{\mathscr{M}}}

\def\cblue{\color{blue}}

\begin{document}

\title{Reverberation Mapping Measurements of Black Hole Masses and Broad-Line Region Kinematics in Mrk~817 and NGC~7469} 

\author{Kai-Xing Lu\altaffilmark{1,2}, 
Jian-Guo Wang\altaffilmark{1,2},  
Zhi-Xiang Zhang\altaffilmark{3}, 
Ying-Ke Huang\altaffilmark{4}, 
Liang Xu\altaffilmark{1,2}, 
Yu-Xin Xin\altaffilmark{1,2}, 
Xiao-Guang Yu\altaffilmark{1,2}, 
Xu Ding\altaffilmark{1,2}, 
De-Qing Wang\altaffilmark{1,2}, 
Hai-Cheng Feng\altaffilmark{1,2,5} 
}

\altaffiltext{1}{Yunnan Observatories, Chinese Academy of Sciences, Kunming 650011, People's Republic of China; {\cblue lukx@ynao.ac.cn}}
\altaffiltext{2}{Key Laboratory for the Structure and Evolution of Celestial Objects, Chinese Academy of Sciences, Kunming 650011, People's Republic of China}
\altaffiltext{3}{Department of Astronomy, Xiamen University, Xiamen, Fujian 361005, People's Republic of China}
\altaffiltext{4}{Qian Xuesen Laboratory of Space Technology, China Academy of Space Technology, 104 Youyi Road, Haidian District, Beijing, 100094, People's Republic of China}
\altaffiltext{5}{University of Chinese Academy of Sciences, Beijing 100049, People’s Republic of China}

\begin{abstract} 
We present the results from a spectroscopic monitoring campaign to obtain reverberation-mapping measurements 
and investigate the broad-line region kinematics for active galactic nuclei (AGN) of Mrk~817 and NGC~7469. 
This campaign was undertaken with the Lijiang 2.4-meter telescope, 
the median spectroscopic sampling is 2.0 days for Mrk~817 and 1.0 days for NGC~7469. 
We detect time lags of the broad emission lines including H$\beta$,
H$\gamma$, He~{\sc ii} and He~{\sc i} for both AGNs, and including Fe~{\sc ii} for Mrk~817 with respect to the varying AGN continuum at 5100~\AA. 
Investigating the relationship between line widths and time lags of the broad emission lines, 
we find that the BLR dynamics of Mrk~817 and NGC~7469 are consistent with the virial prediction. 
We estimate the masses of central supermassive black hole (SMBH) and the accretion rates of both AGNs. 
Using the data of this campaign, 
we construct the velocity-resolved lag profiles of the broad H$\gamma$, H$\beta$, and He~{\sc i} lines for Mrk~817, 
which show almost the same kinematic signatures that the time lags in the red wing are slightly larger than the time lags in the blue wing. 
For NGC~7469, we only clearly construct the velocity-resolved lag profiles of the broad H$\gamma$ and H$\beta$, 
which show very similar kinematic signatures to the BLR of Mrk~817. 
These signatures indicate that the BLR of Keplerian motion in both AGNs seemingly has outflowing components during the monitoring period. 
We discuss the kinematics of the BLR and the measurements including SMBH mass and accretion rates. 
\end{abstract}


\section{Introduction}
\label{sec_intro}
Reverberation mapping (RM; \citealt{Blandford1982,Peterson1993}) 
has been extensively used to measure the mass of accreting supermassive black hole (SMBH) and 
investigate the kinematics of the broad-line region (BLR)  in active galactic nuclei (AGN) through intensively spectroscopic monitoring. 
Up to now, researchers had obtained RM measurements for more than 100 AGNs based on the different RM campaigns 
(e.g., \citealt{Peterson1998,
Kaspi2000,Denney2010,Bentz2009,
Barth2011,Du2014,
Lu2019a,Grier2017,Fausnaugh2017apj,Zhang2019,Huang2019,Brotherton2020,Feng2021,Hu2021}, etc.). 
The canonical Radius$-$Luminosity relationship was constructed from several RM campaigns before 2012 
\citep{Wandel1999,Kaspi2000,Bentz2013}, and discussed in a series of works (e.g., \citealt{Czerny2019,Du2019,Yu2020}), 
which provides an indirect way to estimate SMBH mass from single spectrum across cosmic time. 
However, as the parameter space of AGN properties increases via increasing RM sample or repeating/multi-season RM for the interesting AGNs, 
more and more RM results increasingly deviate from this relation ($\sim$0.3~dex; \citealt{Grier2017,Du2018}). 
\cite{Fonseca2020} found that the deviations depend on ultraviolet (UV)$/$optical SED and the relative amount of ionizing radiation. 
In some degree, this is consistent with the previous finding from NGC~5548 that the BLR size follows the varying optical luminosity at 5100~\AA~(\citealt{Lu2016}), 
but \cite{Lu2016} extra found that the variations of the BLR size lag behind the varying optical luminosity. 
\cite{Lu2019a} also found that the BLR size of Mrk~79 is significantly smaller in the low-luminosity state than normal-luminosity state. 
Like NGC~5548, developing long-term RM campaign for more famous AGNs would help us to understand the variations of BLR with the varying ionizing radiation. 
The kinematic structures of the BLR for $\sim$30 AGNs had been probed using velocity-resolved RM 
(e.g., \citealt{Denney2009,Bentz2009,Grier2013a,DeRosa2018}), 
which usually include virialized disc, inflow, and outflow \citep{Bentz2009,Grier2013a}. 
Repeating velocity-resolved RM for the interesting AGNs in different accreting state provides a chance 
to research the connections between the BLR and accretion disc \citep{Xiao2018}. 

Motivated by above perspectives, we conduct a new spectroscopic monitoring campaign for Mrk~817 and NGC~7469. 
For both targets, researchers had developed multi-wavelength monitoring to research the physical connections between the different bands. 
Based on optical RM campaign, although Mrk~817 and NGC~7469 had been monitored more than one time, 
the velocity-resolved maps of the broad emission lines in both targets are not recovered reliably (\citealt{Peterson1998,Denney2010,Collier1998,Peterson2014}).  

The paper is organized as follows. 
We describe the observation and data reduction in detail in Section 2. 
Data analysis and results are presented in Section~3. 
Section~4 is discussion and summary is given in Section~5. 
We use a cosmology with $H_0=67.8{\rm~km~s^{-1}~Mpc^{-1}}$, $\Omega_{\Lambda}=0.692$, 
and $\Omega_{\rm M}=0.308$ \citep{Planck2016}, 

\section{Observation and Data Reduction}
\label{sec_od}
Long slit spectroscopy and data analysis employed in this work follow closely the previous processes 
laid out by \cite{Lu2019a} for the results from the reverberation campaign on Mrk~79. 
The reader is also referred to previous works, such as \cite{Du2014} and \cite{Lu2016}, 
for more details and discussions on these experiments. 

\subsection{Targets}
\label{sec_target}
Mrk~817 and NGC~7469 are selected in this spectroscopic monitoring campaign. 
The major AGN properties and research results for both targets are summarized below. 

Mrk~817 (alternate name PG~1434+590) is a Seyfert 1.5 galaxy (\citealt{Landt2008})  at $z=0.03145$ (NED), its spectrum shows prominent Fe~{\sc ii} emission lines. 
Mrk~817 had been monitored by four RM campaigns (\citealt{Peterson1998,Denney2010}) before the year of 2018. 
Behind our RM campaign of Mrk~817, 
the AGN STORM 2 campaign has obtained a new H$\beta$ time lag of $23.2\pm1.6$~days with respect to the optical continuum at 5100~\AA~(\citealt{Kara2021arXiv})
\footnote{Because a complete results of optical RM measurement from the AGN STORM 2 campaign will be published in a series works (see \citealt{Kara2021arXiv}), 
the discussion of this work does not include this campaign for the time being.}. 
The velocity-resolved lag profiles of H$\beta$ recovered by \cite{Denney2010} marginally show an outflow signature. 
\cite{Winter2011} found no correlation between the X-ray and ultraviolet$/$optical emissions. 
In recent monitoring period (2017 January 2 to 2018 April 20) of X-ray and ultraviolet bands, 
\cite{Morales2019} also found no significant correlation between the X-ray and ultraviolet emissions at any given lag. 
\cite{Morales2019} offered plausible physical explanations why strong correlation might be absent. For example, 
geometric beaming or a funnel in the inner disc might prevent the reprocessing of X-ray into ultraviolet flux; 
A dense outflow from the inner disc could inhibit the reprocessing of X-ray into ultraviolet flux. 
The latter is the most potential reason because outflowing gases were revealed in Mrk~817 by multi-wavelength spectroscopic observations 
(e.g, \citealt{Ilic2006,Winter2011,Miller2021}). 

NGC~7469 (alternate name Mrk~1514) is a Seyfert 1.5 galaxy (\citealt{Landt2008}) at $z=0.01632$ (NED), 
its spectrum shows somewhat weaker Fe~{\sc ii} emission lines. 
\cite{Shapovalova2017} found that its broad-line region is geometrically stable through 20 years of spectral monitoring, 
such as the H$\beta$ has the full width at half maximum (FWHM) around 2000 km~s$^{-1}$ and show a red asymmetry,  the latter is no changes in the 20-year period. 
\cite{Collier1998} obtained the first RM measurements of NGC~7469, 
the latest RM campaign was developed by \cite{Peterson2014} in 2010 to recover the velocity-resolved maps of broad emission lines, 
but this expectation has not been achieved because of a very low level of variability during that monitoring period. 
NGC~7469 had been studied extensively in the ultraviolet and X-ray bands 
(e.g., \citealt{Wanders1997,Kriss2003,Peretz2018,Pahari2020}). 
These researches found that (1) the ultraviolet emissions well correlate with X-ray, this is different with Mrk~817; 
(2) Two absorbers in the ultraviolet band have outflow velocities of $\sim-600~{\rm km~s^{-1}}$ and $\sim-2000~{\rm km~s^{-1}}$, 
and suggested that the low-velocity component located near the broad emission-line region was a highly ionized and high-density absorber, 
the high-velocity component likely resided farther from the central engine was a lowly ionized and low-density absorber. 
A parallel but more complicated emission-line region was observed by \cite{Grafton-Waters2020}. 

\subsection{Spectroscopic Observation}
\label{sec_obs}
The spectroscopic observation of Mrk~817 and NGC~7469 were taken using 
Yunnan Faint Object Spectrograph and Camera (YFOSC) mounted on the Lijiang 2.4-meter telescope (LJT), 
which locates in Lijiang and is administered by Yunnan Observatories, Chinese Academy of Sciences. 
The detailed information of the observation site including the telescope, 
instruments, observing conditions and some calibration methods are described in \cite{Fan2015}, \cite{Wang2019}, \cite{Xin2020} and \cite{Lu2021}. 

During the long slit spectroscopy, we oriented long slit to take the spectra of AGN and a 
nearby non-varying comparison star simultaneously, this observation method was described 
in detail by \citet{Maoz1990} and \citet{Kaspi2000}. The comparison star as a reference standard 
can provide a high-precision flux calibration (\citealt{Lu2016,Lu2019a}). In addition, 
the spectra of comparison star can be used to correct the telluric absorption lines of the target spectrum (\citealt{Lu2021}). 
In the light of the average seeing $\sim1.5^{\prime\prime}$ of the observation site, 
we fixed the projected slit width $2.5''$. Grism 14 was adopted to monitor Mrk~817, 
which provides dispersion of 1.8~\AA\,~pixel$^{-1}$ and covers the wavelength range 3600$-$7460 \AA\,. 
Grism 3 was adopted to monitor NGC~7469, which provides dispersion of 2.9~\AA\,~pixel$^{-1}$ 
and covers the wavelength range 3400$-$9100 \AA\,. Standard neon and helium lamps were used for wavelength calibration. 
The spectra of Mrk~817 were monitored from 2018 December 20 to 2019 May 29, 
NGC~7469 were monitored from 2019 October 12 to December 31. 
In total, we obtained 70 spectroscopic observations for Mrk~817, spanning a observation period of 190 days, 
the median and mean sampling intervals are 2.00 and 2.75 days, respectively. 
We obtained 52 spectroscopic observations for NGC~7469, spanning a observation period of 80 days, 
the median and mean sampling intervals are 1.00 and 1.56 days, respectively. 

\subsection{Data reduction}
\label{sec_dr}
The two-dimensional spectroscopic images including Mrk~817 and NGC~7469 were reduced using the standard {\tt IRAF} procedures. 
We extracted all spectra using a uniform aperture of 20 pixels (5.7$^{\prime\prime}$), 
and background was determined from two adjacent regions ($+7.4^{\prime\prime}\sim+14^{\prime\prime}$ 
and $-7.4^{\prime\prime}\sim-14^{\prime\prime}$) on both sides of the aperture region. 
It is worth stressing that, a relatively small extraction aperture contributes to reducing the poisson noise of 
sky background and increasing the signal-to-noise ratio (S/N) of spectrum, 
and relatively high S/N ratio is conducive to the multi-component decomposition of spectrum. 
To calibrate spectral flux for both targets, we produced the fiducial spectrum of the comparison star using data from nights with photometric conditions, 
and obtained a wavelength-dependent sensitivity function comparing the star's spectrum to the fiducial spectrum. 
Then this sensitivity function was applied to calibrate the observed spectra of AGN. 
This calibration method was widely used in our previous similar practices (e.g., \citealt{Lu2019a}). 
For each target, we corrected Galactic extinction from the flux-calibration spectra using the extinction map of \cite{Schlegel1998}, 
corrected wavelength shift usually caused by varying seeing and mis-centering using [O~{\sc iii}]~$\lambda5007$ line as wavelength reference, 
and transformed all spectra into the rest frame. These processed spectra are adopted in the next analysis. 

It should be noted that the standard spectral calibration method assumes that 
the [O~{\sc iii}]~$\lambda5007$ flux is constant and use it as an internal flux calibrator, 
which certainly provides accurate flux calibration of spectra when we adopt a relatively broad slit to monitor AGN's spectra (such as 5.0$^{\prime\prime}$; \citealt{Fausnaugh2017}). 
Practically, \cite{Lu2019a} analyzed the different advantages of above two methods in detail and stressed that using the spectrum of comparison star 
to calibrate spectral flux of target is a better choice if we simultaneously observe the spectra of AGN and its comparison star with a relatively narrow slit (such as 2.5$^{\prime\prime}$). 

\section{Data Analysis and Results}
\label{sec_result}

\subsection{Mean and RMS Spectra}
\label{sec_meanrms}
\begin{figure}[ht!]
\centering
\includegraphics[angle=0,width=0.49\textwidth]{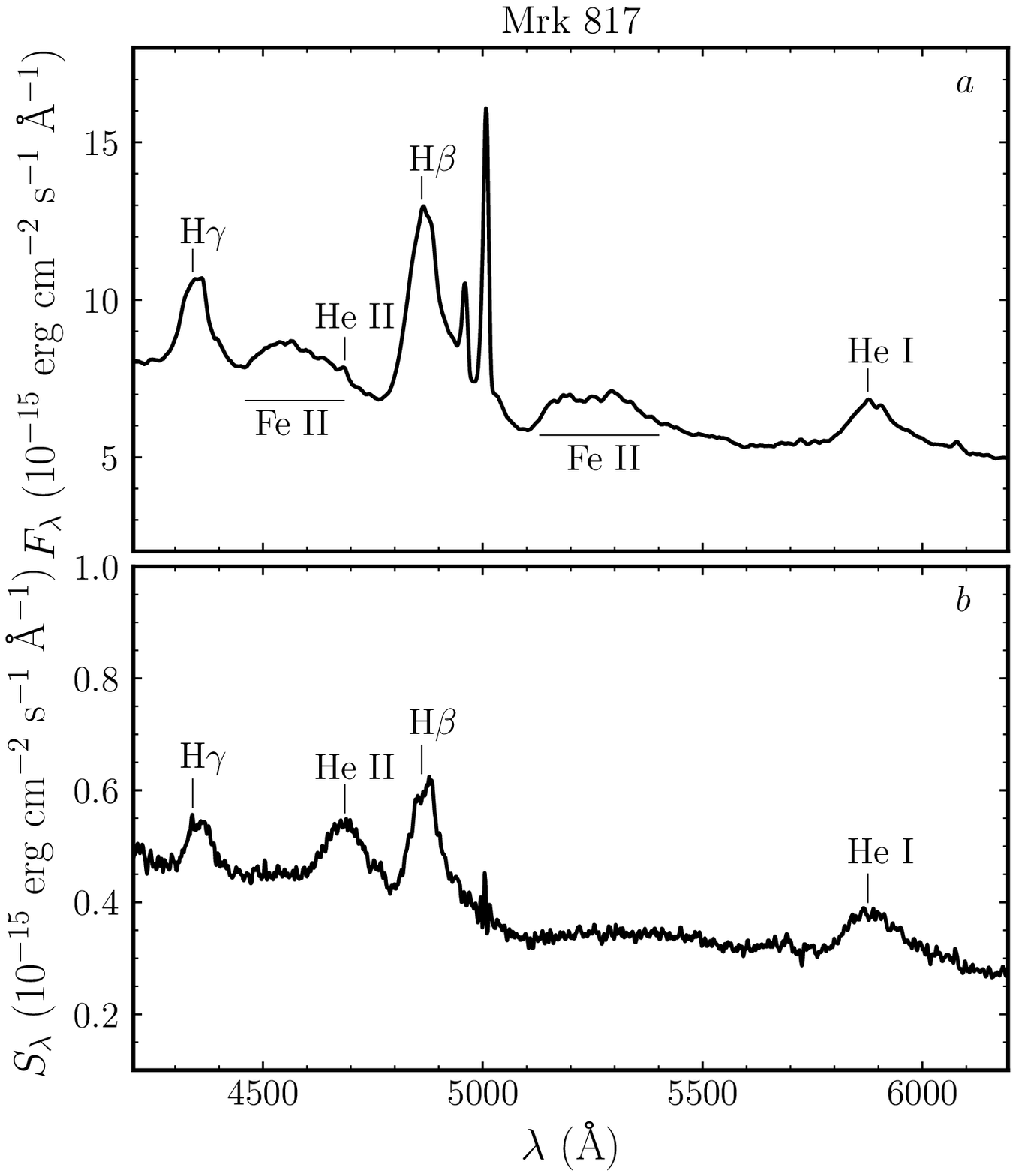}
\includegraphics[angle=0,width=0.49\textwidth]{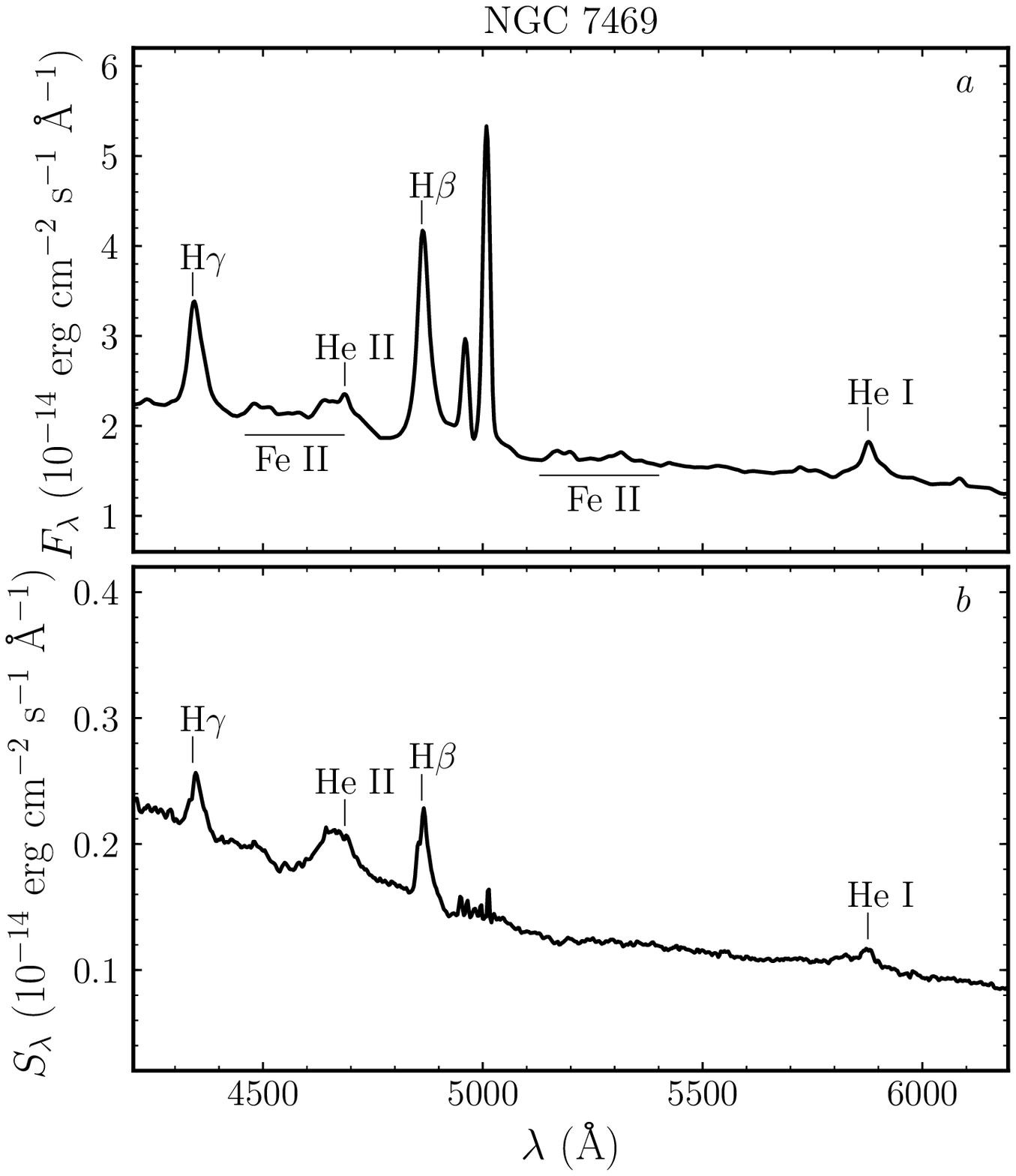}
\caption{\footnotesize
Mean (panel {\it a}) and root mean square spectra (panel {\it b}) of Mrk~817 and NGC~7469 in the rest frame. 
Broad emission lines of Balmer and Helium present significant variability. 
The broad emission lines discussed in the next including H$\gamma$~$\lambda4340$, He~{\sc ii}~$\lambda4686$, H$\beta$~$\lambda4861$, He~{\sc i}~$\lambda5876$ and Fe~{\sc ii} multiplets are labeled.
}
\label{fig_meanrms}
\end{figure}

The definition of the mean spectrum is (\citealt{Peterson2004}) 
\begin{equation}
F_{\lambda}=\frac{1}{N}\sum_{i=1}^NF_i(\lambda), 
\label{eq_msp}
\end{equation}
where $F_i(\lambda)$ is the $i$th spectrum, and $N$ is the total number of spectra obtained in the monitoring period. 
The rms spectrum is 
\begin{equation}
S_{\lambda}=\left\{\frac{1}{N-1}\sum_{i=1}^N\left[F_i(\lambda)-F(\lambda)\right]^2\right\}^{1/2}. 
\label{eq_rsp}
\end{equation}
For each target, we calculate the mean spectrum and root mean square (rms) spectrum from above processed spectra, 
where the [O~{\sc iii}] doublets are subtracted before calculating rms spectrum. 
We presult the results in Figure~\ref{fig_meanrms}. 
Mrk~817 has strong Fe~{\sc ii} emissions while NGC~7469 has relatively weak Fe~{\sc ii} emissions, 
this distinction clearly presents in the mean spectra (panel~{\it a}). 
Broad emission lines of both targets present significant variability in the rms spectra (panel~{\it b}). 

\begin{figure*}[ht!]
\centering
\includegraphics[angle=0,width=0.50\textwidth]{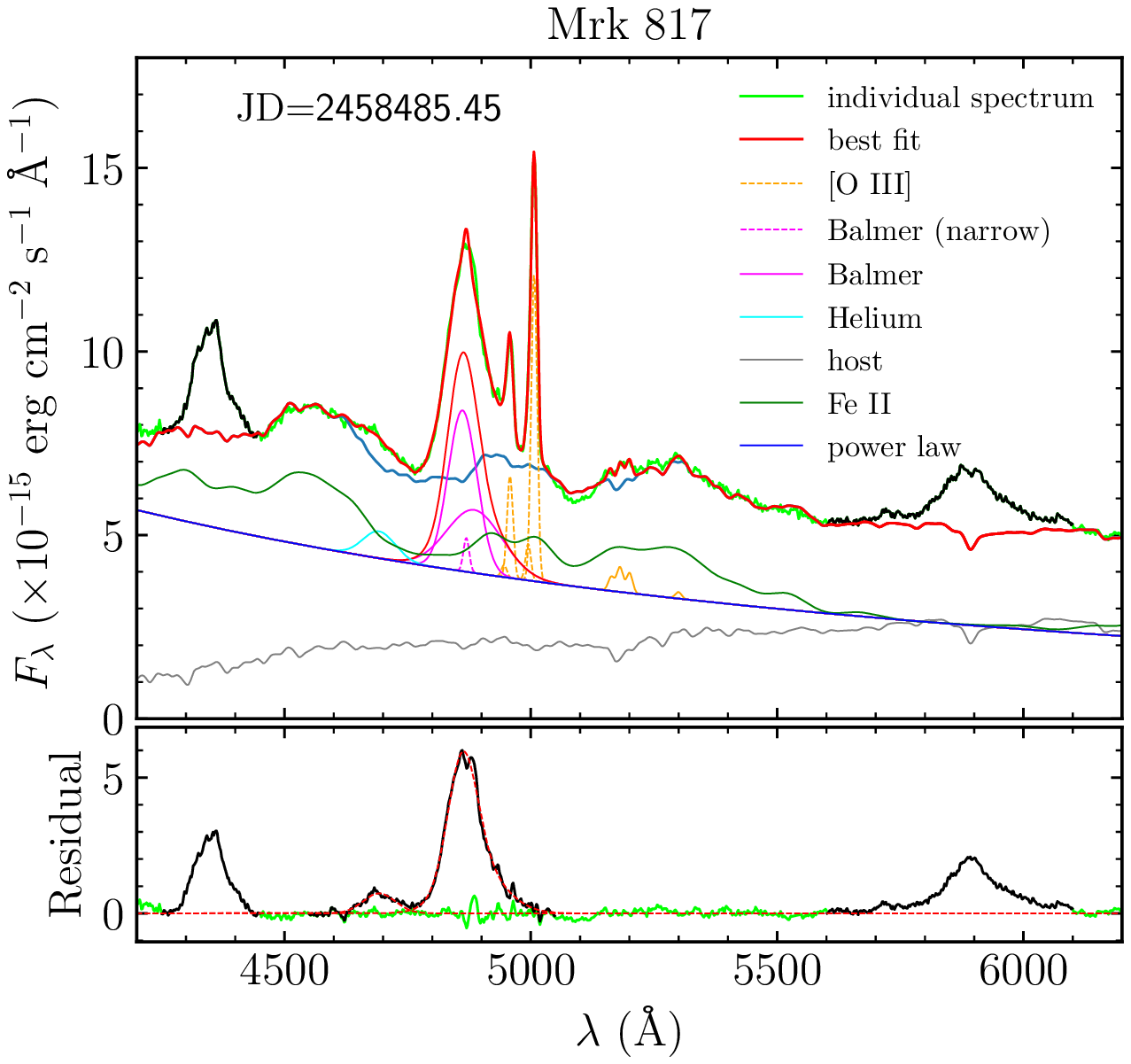}
\includegraphics[angle=0,width=0.49\textwidth]{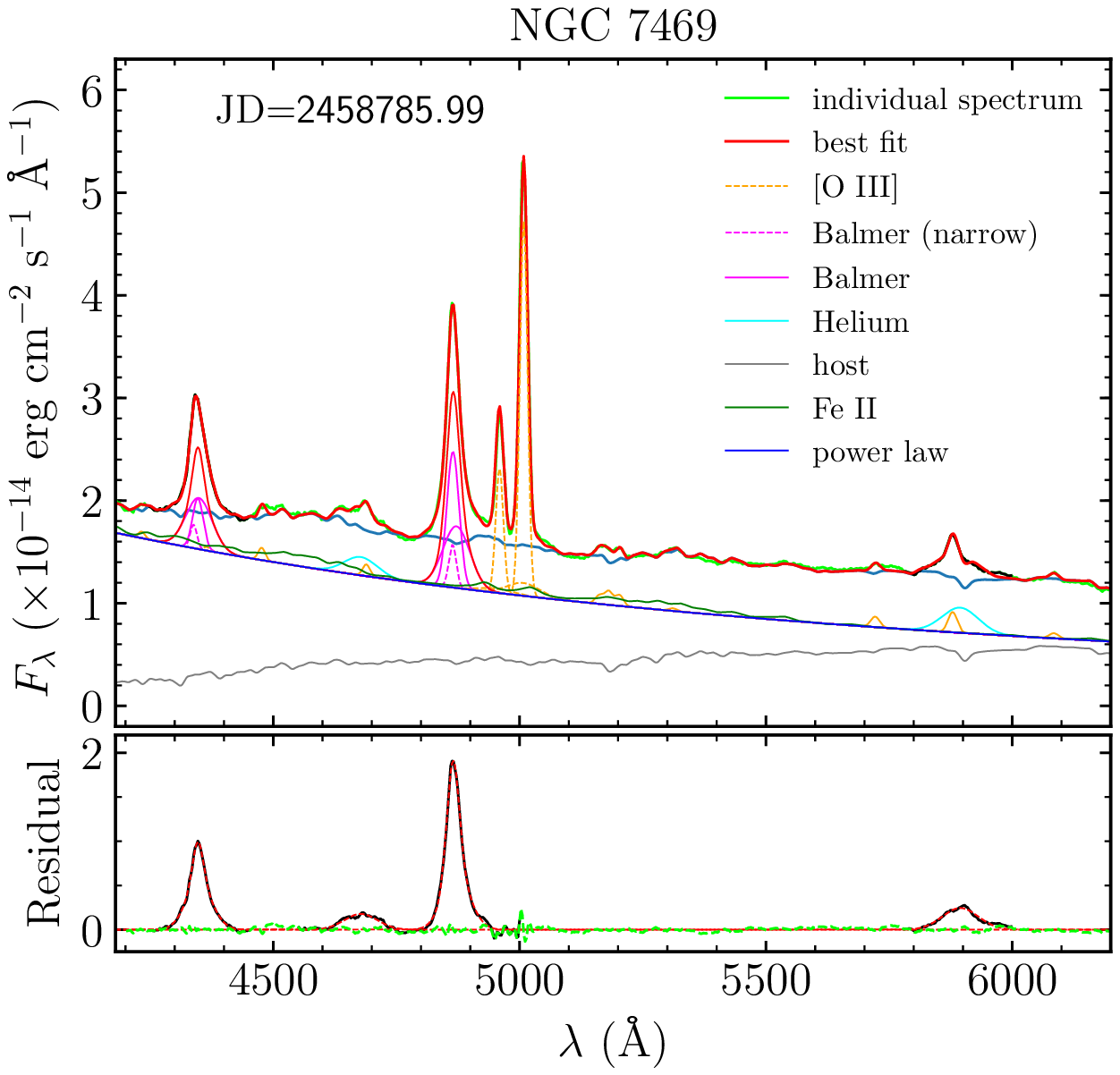}
\caption{\footnotesize
Spectral multi-component decompositions of Mrk~817 (left) and NGC~7469 (right). 
In each case, the top panel shows the details of spectral decompositions, where processed spectrum shows in green (masked region in black), 
those modelled and prominent components present in the legend. 
Some very weak narrow-emission lines including He~{\sc i}~$\lambda4471$, He~{\sc ii}~$\lambda4686$, He~{\sc i}~$\lambda5876$, [N~{\sc i}]~$\lambda5200$, [Fe\,VI] $\lambda5178$, 
[Fe\,VII] $\lambda5158$ and the $\lambda\lambda5721,6086$, and 
[Ca V] $\lambda5310$ are fitted with a set of Gaussian functions and over-plotted in the top panel with the thin solid lines (orange). 
The bottom panel shows the residuals of fitting windows (in green) and net broad lines 
(fitted model in red dashed lines, subtracted case from other fitted components in black). 
}
\label{fig_specfit}
\end{figure*}

\begin{figure*}[ht!]
\centering
\includegraphics[angle=0,width=0.49\textwidth]{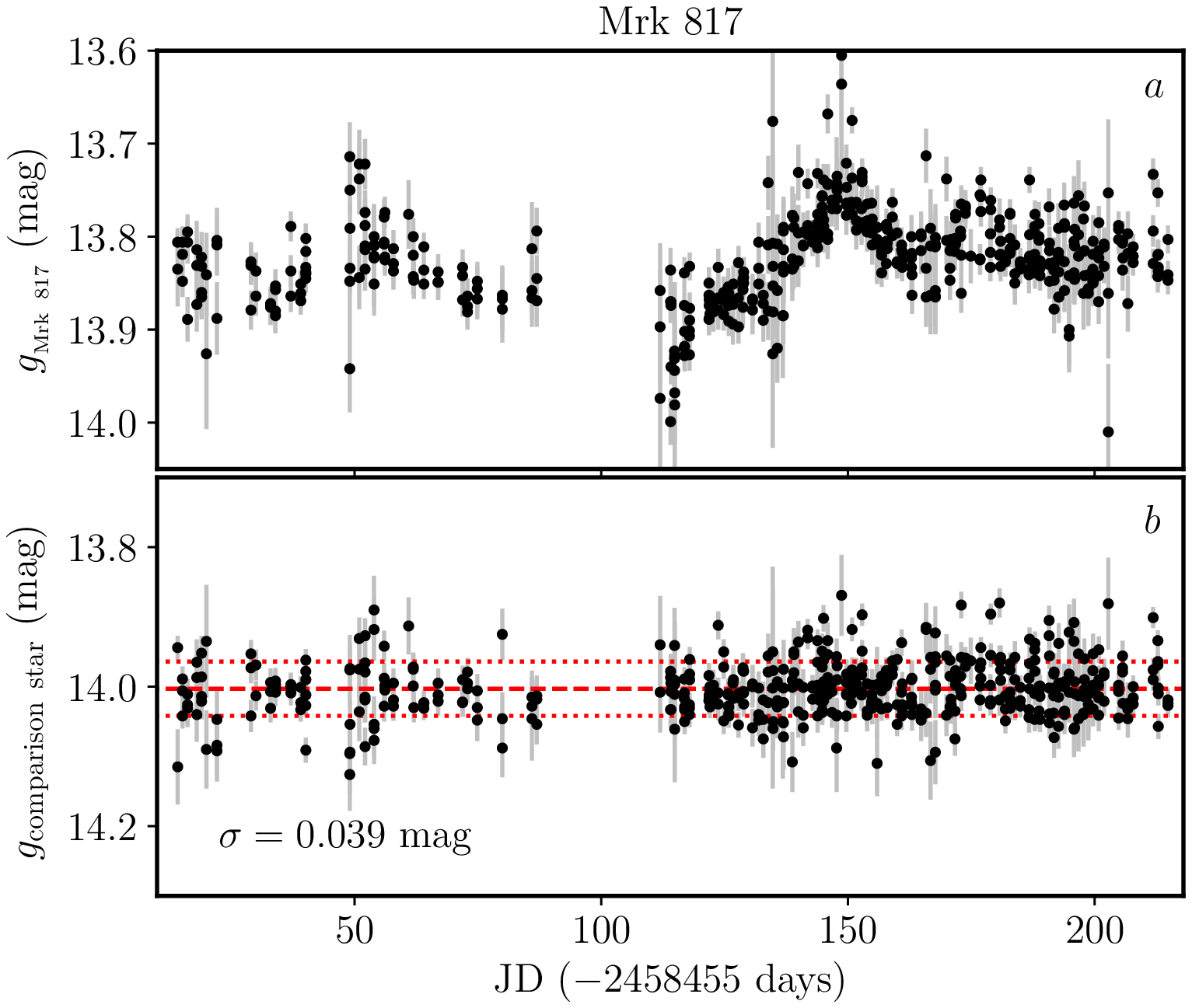}
\includegraphics[angle=0,width=0.50\textwidth]{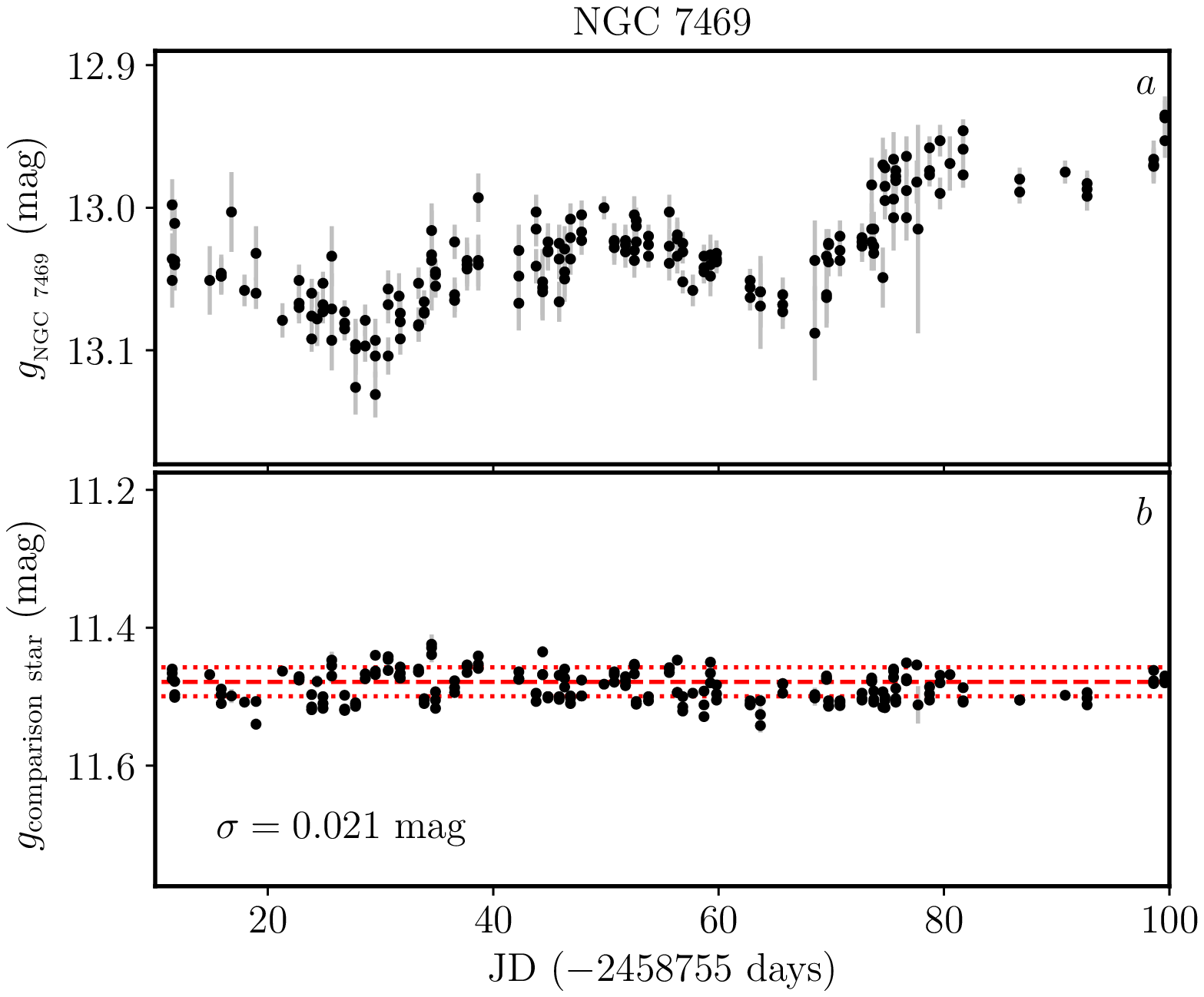}
\caption{\footnotesize
Photometric data of Mrk~817 and NGC~7469 and their comparison star (in g band) compiled from ASAS-SN. In each case, 
the top panel is the light curves of target, the bottom panel is the light curves of its comparison star. 
}
\label{fig_phot}
\end{figure*}

\subsection{Light Curves}
\label{sec_lc}
Spectral fitting method is widely used in the spectroscopic measurements of AGN (e.g., \citealt{Dong2011,Wang2009,Lu2019b}). Especially in RM studies, 
this method is necessary in order to measure the AGN continuum flux without contamination or dilution by these other components.(e.g., \citealt{Barth2013,Hu2015}). 
For example, the starlight from host galaxy dilutes the variability of the AGN continuum and introduces additional random noise due to 
nightly variations in seeing and target centering within the slit (see Figure 11 and 12 of \citealt{Lu2019a}). 
We can improve the measurement quality of light curves of AGN continuum and broad emission lines using spectral fitting and decomposition to 
isolate different portions of the variable spectra (e.g., \citealt{Barth2013,Lu2019a}). 

Following the previous works (e.g., \citealt{Hu2015,Lu2019a}), 
we measure the flux of AGN continuum at 5100~\AA~and broad emission lines (e.g., H$\beta$, H$\gamma$, He~{\sc ii}, He~{\sc i} and Fe~{\sc ii}) 
via a multi-component spectral fitting process. 
The spectra obtained in the present campaign have the second order contamination from the grating at the observed-frame wavelength longer than $\sim$6400~\AA, 
therefore spectral fittings of both targets are performed in the rest-frame wavelength range 4200~\AA$-$6200~\AA~(Figure~\ref{fig_specfit}). 
Because the spectra of both targets have significant different pseudo-continuum 
(composed of the AGN power-law continuum, Fe~{\sc ii} emissions and host galaxy) features 
underneath the broad H$\beta$, H$\gamma$ and He~{\sc i} emission lines (see panel~{\it a} of Figure~\ref{fig_meanrms}), 
we adopt slightly different  fitting procedures for Mrk~817 and NGC~7469. 
Specifically, the broad H$\gamma$ and He~{\sc i} emission lines of Mrk~817 are blended mightily with 
the strong Fe~{\sc ii} emissions result in they are hard to be fitted steadily, 
we mask the H$\gamma$ and He~{\sc i} regions during the fitting of Mrk~817 (left panels of Figure~\ref{fig_specfit}); 
The shift of the broad He~{\sc ii} emission line in Mrk~817 is fixed because it's too weak to restrict the model, 
while the shift of the broad He~{\sc ii} emission line in NGC~7469 is variable. 
The flux ratio of [O~{\sc iii}] doublets is fixed to the theoretical value of 3. 
We find that the broad He~{\sc ii} emission line of NGC~7469 has a mean blue-shifted velocity of $832\pm192~{\rm km~s^{-1}}$. 
We fix the spectral index of individual spectrum to the value given by the best fit of the mean spectrum for both targets, respectively. 
The rest of model parameters are allowed to vary. 
The fitting components include: 
(1) a single power law ($f_{\lambda}\propto\lambda^{\alpha}$, $\alpha$ is the spectral index) for the AGN continuum. 
(2) Fe~{\sc ii} multiplets from \cite{Boroson1992}; 
(3) the template of host galaxy with 11 Gyr age and metallicity Z = 0.05 from \cite{Bruzual2003}; 
(4) Gaussian function for the broad-emission lines; 
(5) two double Gaussians for the [O~{\sc III}] doublets $\lambda5007/\lambda$4959; 
(6) a set of single Gaussians with the same velocity and shift for considering narrow emission lines. 
We fit the above components simultaneously to the spectra of both targets in the fitting region. 

Figure~\ref{fig_specfit} shows the multi-component fitting results of individual spectrum for Mrk~817 and NGC~7469. 
In each case, we measure AGN continuum flux at 5100~\AA~($F_{\rm 5100}$) from the best fit of the power-law component. 
In the bottom panels of Figure~\ref{fig_specfit}, in addition to showing the residuals for each case, 
we also show the components of net broad emission lines including H$\beta$, H$\gamma$, He~{\sc ii} and He~{\sc i}, 
where the fitted model shows in red dashed lines and the subtracted case by eliminating other fitted components shows in black, 
the latter is adopted in the next RM measurements. 
We measure the flux of this broad emission lines from the subtracted case. 
Following \cite{Boroson1992}, we measure the flux of the integrated Fe~{\sc ii} ($F_{\rm Fe}$) 
emission between 4434~\AA~and 4684~\AA~from the best-fit Fe~{\sc ii} model. 
Because Fe~{\sc ii} emissions of NGC~7469 are weak, we do not measure its flux. 
Table~\ref{tab1}~and~\ref{tab2} summarise the light curves of Mrk~817 and NGC~7469, 
where the errors of the light curves include the Poisson errors and additional systematic errors (which are calculated via median filter). 
It should be noted that the previous works found the contamination of host galaxy to AGN continuum can be will estimated and eliminated 
by spectral fitting and decomposition (e.g., \citealt{Hu2015,Lu2019a}), therefore above measured flux including AGN continuum at 5100~\AA~and 
all broad emission lines have eliminated the contamination of host galaxy. 

During our spectroscopic monitoring periods, both targets and their comparison stars were also monitored using g band filter by 
the project of All-Sky Automated Survey for Super-novae\footnote{http://www.astronomy.ohio-state.edu/asassn/index.shtml} 
(ASAS-SN; \citealt{Shappee2014,Kochanek2017}). 
We compile photometric data of two targets and their comparison stars from archived data of ASAS-SN 
and plot them in Figure~\ref{fig_phot}. 
The light curves including AGN continuum at 5100~\AA~($F_{\rm 5100}$) and broad emission lines measured in the present campaign 
are plotted in the left panels of Figure~\ref{fig_mrkmap}~and~\ref{fig_ngcmap} for Mrk~817 and NGC~7469, respectively. 
Inspecting the light curves of AGN continuum at 5100~\AA~obtained from our spectroscopy 
($F_{\rm 5100}$, panel~{\it a} of Figure~\ref{fig_mrkmap}~and~\ref{fig_ngcmap}) 
with the photometric light curves in g band from ASAS-SN (panel~{\it a} Figure~\ref{fig_phot}) for Mrk~817 and NGC~7469, respectively, 
we find that spectroscopic data and photometric data present the similar structure of variability for each target. 
The light curves of Mrk~817's and NGC~7469's comparison stars from ASAS-SN have the scatter of 3.9\% and 2.1\%, respectively. 
Under the poor photometric accuracy, both stars are still stable enough, 
which demonstrate that using both stars as the spectroscopic comparison star of Mrk~817 and NGC~7469 are robust. 

\begin{deluxetable*}{cccccccc}
\tablecolumns{12}
\tabletypesize{\scriptsize}
\tablewidth{0pt}
\tablecaption{Light curves of AGN continuum at 5100~\AA~and broad emission lines for Mrk~817\label{tab1}}
\tablehead{
\colhead{JD$-$2458455}                &
\colhead{$F_{\rm 5100}$}                &
\colhead{$F_{\rm H\beta}$}             &
\colhead{$F_{\rm H\gamma}$}        &
\colhead{$F_{\rm He~{\sc II}}$}       &
\colhead{$F_{\rm He~{\sc I}}$}   &
\colhead{$F_{\rm Fe}$}   
}
\startdata
18.461 &$ 3.683 \pm 0.077 $&$ 5.146 \pm 0.087 $&$ 2.031 \pm 0.039 $&$ 7.155 \pm 0.185 $&$ 1.947 \pm 0.031 $&$ 4.166 \pm 0.069 $\\
19.443 &$ 4.167 \pm 0.077 $&$ 5.677 \pm 0.087 $&$ 2.212 \pm 0.043 $&$ 8.107 \pm 0.222 $&$ 2.110 \pm 0.031 $&$ 4.794 \pm 0.069 $\\
21.468 &$ 3.523 \pm 0.116 $&$ 5.174 \pm 0.088 $&$ 2.200 \pm 0.045 $&$ 6.632 \pm 0.282 $&$ 1.943 \pm 0.033 $&$ 4.152 \pm 0.071 $\\
22.464 &$ 3.523 \pm 0.107 $&$ 5.162 \pm 0.088 $&$ 2.050 \pm 0.043 $&$ 4.545 \pm 0.258 $&$ 1.893 \pm 0.033 $&$ 3.825 \pm 0.070 $
\enddata
\tablecomments{\footnotesize
$F_{\rm 5100}$ is in units of ${\rm 10^{-15}~erg~s^{-1}~cm^{-2}~\AA^{-1}}$. 
$F_{\rm H\gamma}$, $F_{\rm H\beta}$, $F_{\rm He~{\sc I}}$ and $F_{\rm Fe}$ are in the units of ${\rm 10^{-13}~erg~s^{-1}~cm^{-2}}$. 
$F_{\rm He~{\sc II}}$ is in units of ${\rm 10^{-14}~erg~s^{-1}~cm^{-2}}$. 
This table is available in its entirety in machine-readable form.
}
\end{deluxetable*}

\begin{deluxetable*}{ccccccc}
\tablecolumns{11}
\tabletypesize{\scriptsize}
\tablewidth{0pt}
\tablecaption{Light curves of AGN continuum at 5100~\AA~and broad emission lines for NGC~7469\label{tab2}}
\tablehead{
\colhead{JD$-$2458755}                &
\colhead{$F_{\rm 5100}$}                &
\colhead{$F_{\rm H\beta}$}             &
\colhead{$F_{\rm H\gamma}$}        &
\colhead{$F_{\rm He~{\sc II}}$}       &
\colhead{$F_{\rm He~{\sc I}}$}     
}
\startdata
14.099 &$ 1.163 \pm 0.022 $&$ 8.982 \pm 0.067 $&$ 5.530 \pm 0.078 $&$ 3.494 \pm 0.060 $&$ 2.311 \pm 0.026 $\\
15.187 &$ 1.233 \pm 0.022 $&$ 8.884 \pm 0.068 $&$ 5.447 \pm 0.083 $&$ 2.511 \pm 0.061 $&$ 2.336 \pm 0.025 $\\
19.211 &$ 1.077 \pm 0.022 $&$ 8.680 \pm 0.067 $&$ 4.953 \pm 0.079 $&$ 2.471 \pm 0.060 $&$ 2.178 \pm 0.025 $\\
21.195 &$ 1.038 \pm 0.021 $&$ 8.598 \pm 0.066 $&$ 4.889 \pm 0.076 $&$ 2.288 \pm 0.058 $&$ 2.176 \pm 0.024 $
\enddata
\tablecomments{\footnotesize
$F_{\rm 5100}$ is in units of ${\rm 10^{-14}~erg~s^{-1}~cm^{-2}~\AA^{-1}}$. 
All broad emission lines in this table are in units of  ${\rm 10^{-13}~erg~s^{-1}~cm^{-2}}$. 
This table is available in its entirety in machine-readable form.
}
\end{deluxetable*}

\subsection{Variability Characteristics}
\label{sec_vari}
We calculate the variability amplitude of all light curves using equation of (\citealt{Rodriguez-Pascual1997}) 
\begin{equation}
F_{\rm var}=\frac{\left(\sigma^2-\Delta^2\right)^{1/2}}{\langle F\rangle}, 
\label{eq_fvar}
\end{equation}
where $\langle F\rangle$ is the mean flux, $\sigma^2$ is the variance, and $\Delta^2$ is the mean square error. 
The uncertainty of $F_{\rm var}$ is defined as (\citealt{Edelson2002}) 
\begin{equation}
\sigma_{_{F_{\rm var}}} = \frac {1} {F_{\rm var}} \left(\frac {1}{2 N}\right)^{1/2} \frac {\sigma^2}{\langle F\rangle^2}, 
\end{equation}
where $N$ is the total epochs. Table~\ref{tab_lcst} lists the statistics of the light curves. 

\begin{deluxetable}{llcc}
\tablecolumns{4}
\tabletypesize{\scriptsize}
\tablewidth{0pt}
\tablecaption{Light curve statistics\label{tab_lcst}}
\tablehead{
\colhead{Object}                 &
\colhead{Light curves}       &
\colhead{Mean flux}           &
\colhead{$F_{\rm var}$ (\%)} 
}
\startdata
Mrk~817     & $F_{\rm 5100}$                &$ 3.77\pm0.30 $& $7.82\pm0.71$  \\
	          &  $F_{\rm H\beta}$             &$ 5.09\pm0.28 $& $5.36\pm0.50$  \\
	          &  $F_{\rm H\gamma}$        &$1.97\pm0.15 $&  $7.32\pm0.66$ \\
	          &  $F_{\rm He~{\sc II}}$       &$ 0.46\pm 0.21 $&  $46.52\pm3.96$  \\
	          &  $F_{\rm He~{\sc I}}$        &$ 1.91\pm0.12 $&  $5.93\pm0.54$ \\
	          &  $F_{\rm Fe~{\sc II}}$       &$ 3.96\pm0.24 $& $5.81\pm0.53$  \\\hline
NGC~7469 & $F_{\rm 5100}$                &$ 1.24\pm0.15 $& $11.86\pm1.19$ \\
	          &  $F_{\rm H\beta}$            &$ 8.47\pm0.31$& $3.57\pm0.37$ \\
	          &  $F_{\rm H\gamma}$       &$ 4.96\pm0.26 $&  $5.01\pm0.54$ \\
	          &  $F_{\rm He~{\sc II}}$      &$  2.67\pm0.59$& $22.29\pm2.21$ \\
	          &  $F_{\rm He~{\sc I}}$        &$  2.21\pm0.12$&  $5.32\pm0.55$ 
\enddata
\tablecomments{\footnotesize
The $F_{\rm 5100}$ of Mrk~817 is in units of ${\rm 10^{-15}~erg~s^{-1}~cm^{-2}~\AA^{-1}}$, 
the $F_{\rm 5100}$ of NGC~7469 is in units of ${\rm 10^{-14}~erg~s^{-1}~cm^{-2}~\AA^{-1}}$, 
the fluxes of all emission lines are in units of ${\rm 10^{-13}~erg~s^{-1}~cm^{-2}}$.
}
\end{deluxetable}

\begin{deluxetable}{lccccc}
  \tablecolumns{6}
  \tabletypesize{\scriptsize}
  \setlength{\tabcolsep}{3pt}
  \tablewidth{4pt}
  \tablecaption{Time lags in the rest frame of the present campaign\label{tab_rm1}}
  \tablehead{
  \colhead{} &
  \colhead{} &
  \multicolumn{3}{c}{Line vs.~$F_{\rm 5100}$} &
  \colhead{}  \\ \cline{3-5} 
  \colhead{Object}&
  \colhead{Lines}&
  \colhead{$\tau_{\rm cent}$}&
  \colhead{$\tau_{\rm peak}$}&
  \colhead{$r_{\rm max}$}
  }
\startdata
Mrk~817     &H$\beta$      & $28.3^{+2.1}_{-1.8}$   &$29.5^{+1.2}_{-4.1}$    &  0.78  \\   
                   &H$\gamma$ & $26.8^{+2.8}_{-2.5}$    &$26.6^{+3.5}_{-5.3}$   &  0.59  \\   
                   &He~{\sc ii}    & $13.1^{+3.1}_{-5.6}$    &$6.1^{+15.8}_{-0.6}$   &  0.42   \\   
                   &He~{\sc i}     & $18.6^{+3.8}_{-1.9}$    &$20.8^{+5.3}_{-8.2}$   &  0.67 \\
                   &Fe~{\sc ii}     & $51.7^{+14.9}_{-1.3}$ &$52.4^{+19.3}_{-2.4}$  &  0.48 \\\hline
NGC~7469 &H$\beta$      & $8.0^{+0.8}_{-1.5}$     &$7.3^{+1.1}_{-1.7}$      &  0.88  \\   
                   &H$\gamma$ & $4.2^{+2.1}_{-1.2}$     &$4.6^{+0.8}_{-1.9}$     &   0.83  \\   
                   &He~{\sc ii}    & $1.1^{+1.2}_{-1.0}$      &$0.8^{+1.8}_{-0.9}$     &   0.61  \\   
                   &He~{\sc i}     & $5.7^{+1.5}_{-1.8}$     &$4.5^{+5.0}_{-1.3}$      &   0.68 
\enddata
\tablecomments{\footnotesize
Time lags of $\tau_{\rm cent}$ and $\tau_{\rm peak}$ are measured from the centroid and peak of ICCF, 
$r_{\rm max}$ is the maximum cross-correlation coefficient. 
}
\end{deluxetable}

\subsection{Time Lags of the Broad Emission Lines}
\label{sec_lag}
\begin{figure*}[ht!]
\centering
\includegraphics[angle=0,width=0.8\textwidth]{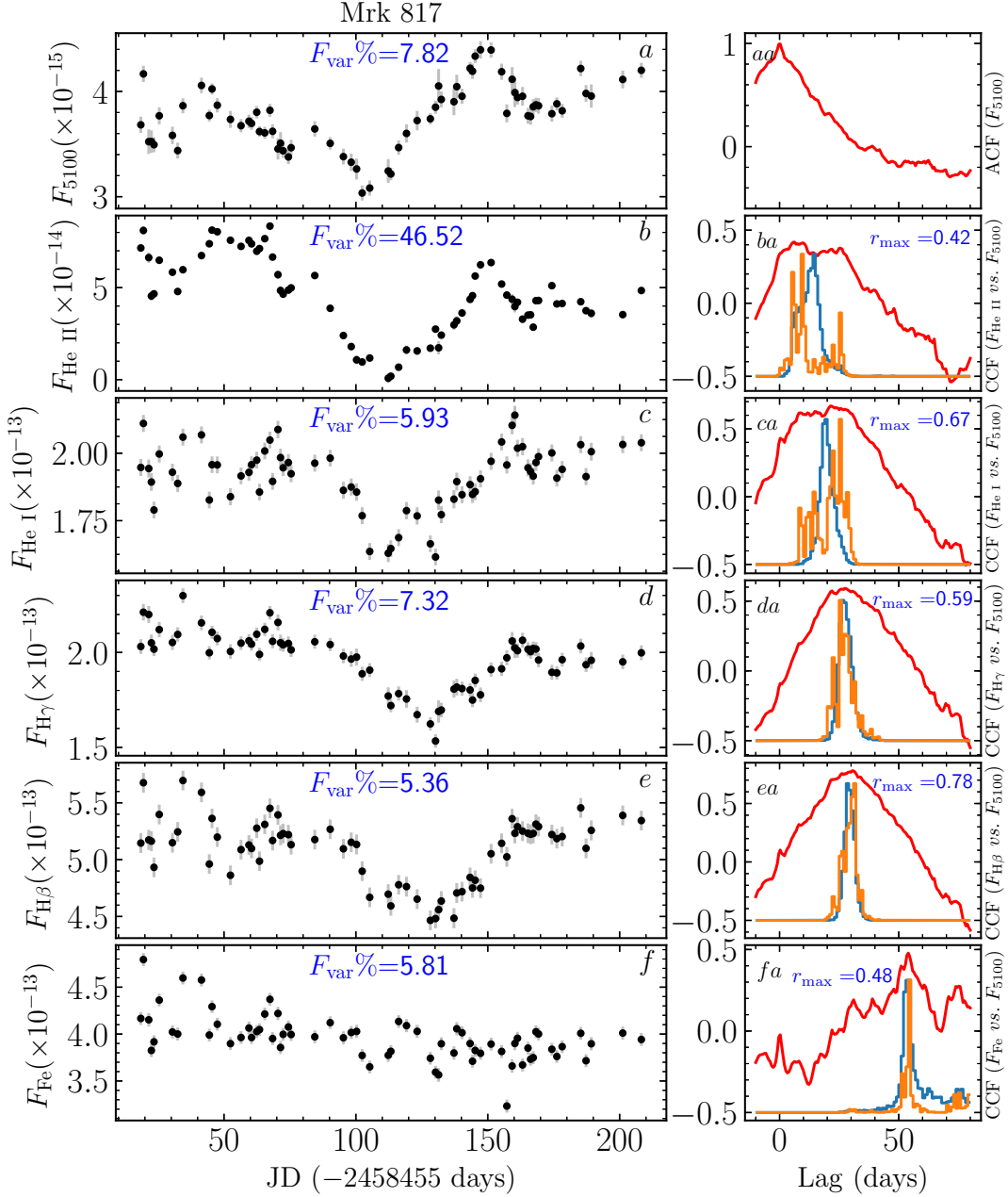}
\caption{\footnotesize
Light curves and the results of cross correlation analysis for Mrk~817.
The left panels ({\it a-f}) are the light curves of AGN continuum at 5100~\AA~and the broad He~{\sc ii}, He~{\sc i}, H$\gamma$, H$\beta$ and Fe~{\sc ii} lines. 
The right panels ({\it aa-fa}) correspond to the ACF of AGN continuum and the CCF between the broad emission lines ({\it b-e}) and the varying AGN continuum ({\it a}), respectively. 
In panels ({\it ba-fa}), we plot the ICCF in red, the CCCD in blue, and the CCPD in orange. 
We note the variability amplitude $F_{\rm var}\%$ in the panels of light curves, 
and note the maximum cross-correlation coefficient ($r_{\max}$) in the panels of CCF. 
The units of $F_{\rm 5100}$ and emission lines are ${\rm erg~s^{-1}~cm^{-2}~\AA^{-1}}$ and ${\rm erg~s^{-1}~cm^{-2}}$, respectively. 
}
\label{fig_mrkmap}
\end{figure*}

\begin{figure*}[ht!]
\centering
\includegraphics[angle=0,width=0.8\textwidth]{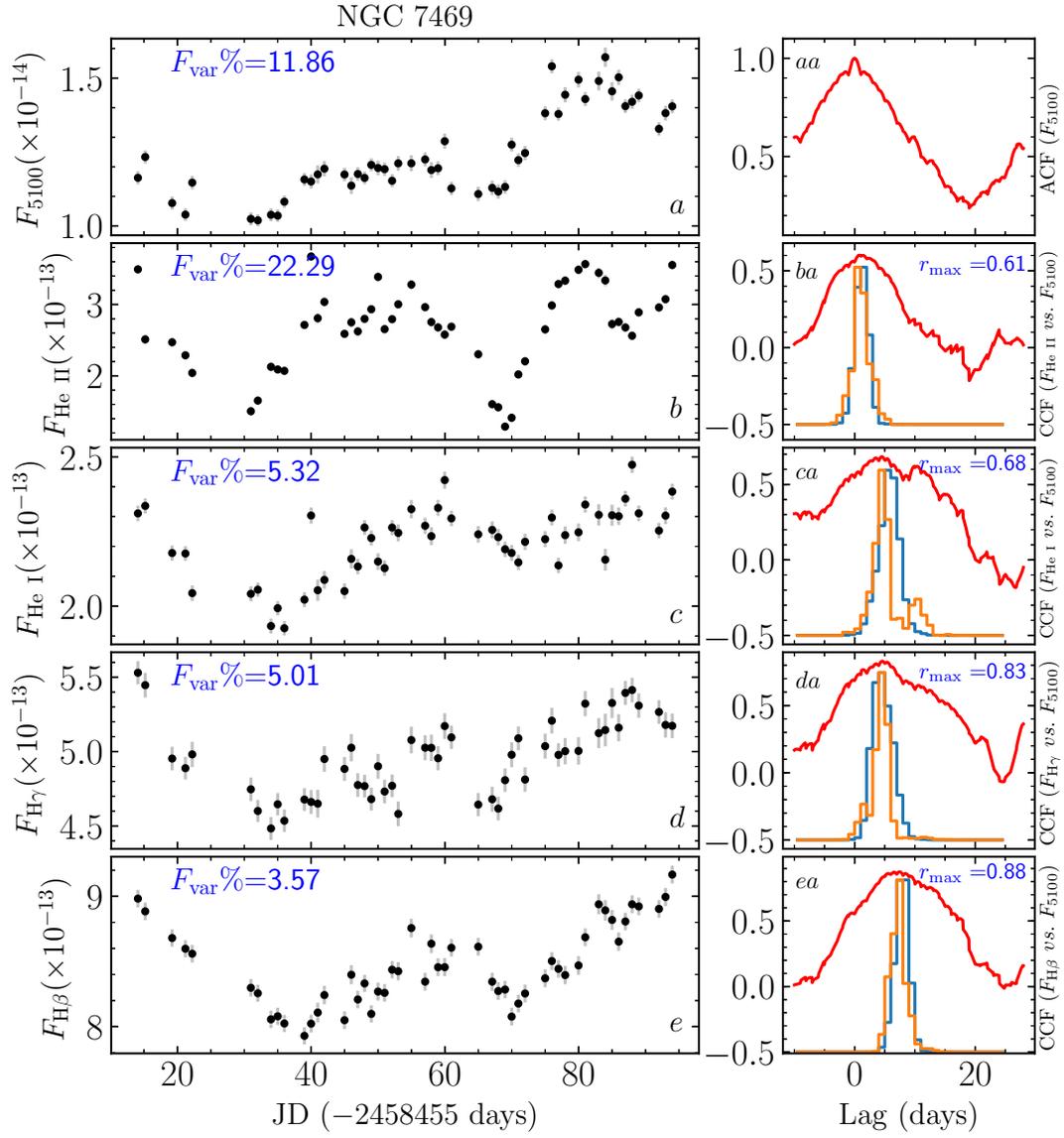}
\caption{\footnotesize
Same as Figure~\ref{fig_mrkmap}, but for NGC~7469.
}
\label{fig_ngcmap}
\end{figure*}

We calculate time lags between the broad emission lines (including H$\beta$, H$\gamma$, He~{\sc ii}, He~{\sc i} for both targets, and Fe~{\sc ii} for Mrk~817) 
and the varying AGN continuum at 5100~\AA~($F_{\rm 5100}$) using interpolation cross-correlation function (ICCF; \citealt{Gaskell1986,Gaskell1987}). 
The locations of ICCF peak and centroid are usually adopted as the time lag, and recorded as $\tau_{\rm peak}$ and $\tau_{\rm cent}$, respectively. 
Using the model-independent ``flux randomization/random subset sampling (FR/RSS)'' method \citep{Peterson1998}, 
we estimate the uncertainties of $\tau_{\rm peak}$ and $\tau_{\rm cent}$ in the 15.87\% and 84.13\% quantiles 
of the cross-correlation centroid and peak distributions(CCCD and CCPD), respectively. 
The cross-correlation analysis results (ICCF, CCCD and CCPD) 
are plotted in the right panels of Figure~\ref{fig_mrkmap}~and~\ref{fig_ngcmap} for Mrk~817 and NGC~7469, respectively. 
The variability amplitudes and the maximum cross-correlation coefficients are noted in the corresponding panels. 
We summarize these measurements in Table~\ref{tab_rm1}. For two targets, except for the weak He~{\sc ii} emission line of Mrk~817, 
the time lag of other broad emission lines measured from the ICCF peak ($\tau_{\rm peak}$) are consistent with corresponding time lag measured from ICCF centroid ($\tau_{\rm cent}$). 
The time lags of these broad emission lines show a stratified BLR for each target, this is consistent with previous findings 
(e.g., \citealt{Bentz2010b,Lu2019a}). 
For Mrk~817, we measure time lag between Fe~{\sc ii} and the varying AGN continuum at 5100~\AA~with the maximum cross-correlation coefficient of $0.48$, 
and find that the time lag of Fe~{\sc ii} is 1.8~times of the H$\beta$ time lag. 
This is consistent with the previous findings (e.g., \citealt{Barth2013,Hu2015}). 

\clearpage

\begin{deluxetable*}{lccccccccc}
  \tablecolumns{10}
  \tabletypesize{\scriptsize}
  \setlength{\tabcolsep}{3pt}
  \tablewidth{4pt}
  \tablecaption{Line widths and time lags (in the rest frame) of Mrk~817 and NGC~7469 from all the available RM campaigns\label{tab_sum}}
  \tablehead{
  \colhead{} &
  \colhead{} &
  \multicolumn{2}{c}{Mean spectra} &
  \colhead{} &
  \multicolumn{2}{c}{rms spectra}  &
   \colhead{} &
  \colhead{}  \\ \cline{3-4} \cline{6-7}  
  \colhead{Object}&
   \colhead{Lines}&
  \colhead{FWHM~(km~s$^{-1}$)}&
  \colhead{$\sigma_{\rm line}$~(km~s$^{-1}$)}&
  \colhead{} &
  \colhead{FWHM~(km~s$^{-1}$)}&
  \colhead{$\sigma_{\rm line}$~(km~s$^{-1}$)} &
   \colhead{} &
   \colhead{$\tau~{\rm (days)}$}  &
    \colhead{References} \\
  \colhead{(1)} &
  \colhead{(2)} &
  \colhead{(3)} &
  \colhead{(4)} &
  \colhead{} &
   \colhead{(6)} &
  \colhead{(7)} &
   \colhead{}  &
  \colhead{(8)} &
   \colhead{(9)} 
 }
\startdata
\multicolumn{9}{c}{Present campaign}\\ \cline{1-10}  
Mrk~817     &H$\beta$      & $4447\pm9$     & $2522\pm7$   && $4464\pm160$ & $2143\pm56$   && $28.32^{+2.12}_{-1.77}$     & 1\\   
                   &H$\gamma$ & $4650\pm23$    & $2447\pm16$ && $5209\pm659$ & $2338\pm298$ && $26.82^{+2.84}_{-2.51}$    & 1\\   
                   &He~{\sc ii}    & $5235\pm70$    & $2207\pm40$ && $6662\pm411$ & $3118\pm173$    && $13.13^{+3.13}_{-5.59}$  & 1  \\   
                   &He~{\sc i}     & $5222\pm31$    & $2849\pm8$   && $4546\pm749$ & $2506\pm76$    && $18.55^{+3.79}_{-1.91}$   & 1 \\
                   &Fe~{\sc ii}     &         ...               & $1268\pm79$ &&             ...           &             ...          && $51.69^{+14.94}_{-1.32}$ & 1  \\
                   \hline
NGC~7469 &H$\beta$      & $1893\pm10$   & $738\pm6$      && $1562\pm99$   & $1485\pm34$   && $8.03^{+0.79}_{-1.46}$    & 1  \\   
                   &H$\gamma$ & $2100\pm19$   & $1214\pm6$    && $2491\pm167$ & $1808\pm55$   && $4.16^{+2.12}_{-1.16}$    & 1  \\   
                   &He~{\sc ii}    & $5468\pm26$   & $2103\pm13$  && $5961\pm183$ & $3272\pm179$  && $1.05^{+1.15}_{-1.00}$    & 1  \\   
                   &He~{\sc i}     & $3994\pm37$   & $1417\pm22$  && $4528\pm198$ & $1609\pm155$   && $5.66^{+1.54}_{-1.76}$  & 1   \\  
 		\hline
 \multicolumn{9}{c}{Previous campaigns}\\ \cline{1-10}  
Mrk~817     &H$\beta$      & $4711\pm49$   &$1984\pm7$     && $3514\pm393$  & $1392\pm78$   && $19.0^{+3.9}_{-3.7}$  & 2,3,4 \\
		  &H$\beta$      & $5236\pm66$   &$2097\pm12$   && $4952\pm537$  & $1970\pm95$   && $15.3^{+3.7}_{-3.5}$  & 2,3,4 \\
		  &H$\beta$      & $4766\pm72$   &$2195\pm16 $  && $3751\pm994$  & $1729\pm158$ && $33.6^{+6.5}_{-7.6}$  & 2,3,4 \\
		  &H$\beta$      &           ...            &            ...          && $5627\pm30$    & $ 2025\pm5$    && $14.0^{+3.4}_{-3.5}$  & 4,5 \\
		  \hline
NGC~7469 &H$\beta$      & $1478\pm21^a$   & $1095\pm5$    && $1066\pm84$    & $1274\pm126$ && $10.8^{+3.4}_{-1.3}$  & 6 \\
		  &He~{\sc ii}    & $2197\pm339$  &$2306\pm8$    &&  $5607\pm315$  & $2271\pm77$  && $1.3^{+0.9}_{-0.7}$     & 6 \\
&C~{\sc iv}~$\lambda1549$ & $1722\pm30$& $1707\pm20$ && $4305\pm422$   & $2619\pm118$ && $2.5^{+0.3}_{-0.2}$    & 6 \\
&Si~{\sc iv}~$\lambda1400$$^{b}$&         ...  &            ...          &&          ...              & $3492\pm34$   && $1.7^{+0.3}_{-0.3}$    & 6 \\
&He~{\sc ii}~$\lambda1640$$^{b}$&          ... &            ...          &&          ...               & $3718\pm37$   && $0.6^{+0.3}_{-0.4}$    & 6 \\
		  &H$\beta$       &$1722\pm29$&$1706\pm19$   &&$2169\pm459$   &$1456\pm207$   && $4.5^{+0.7}_{-0.8}$    & 3,6 \\
		  &H$\alpha$$^{b}$    &          ...     &            ...          &&          ...               & $1162\pm11$    && $4.7^{+1.6}_{-1.3}$    & 6 
\enddata
\tablecomments{\footnotesize
In addition to listing the data of present campaign, we also list the available data of previous campaigns. 
To distinguish optical and ultraviolet (UV) emission lines, we give the information of wavelength for UV emission lines 
(i.e., He~{\sc ii}~$\lambda1640$, C~{\sc iv}~$\lambda1549$ and Si~{\sc iv}~$\lambda1400$) in the Col~(2), 
and other lines without the information of wavelength mean optical emission lines. All time lags ( Col.~8) were obtained from the centroid of CCF.  \\
``$a$" This FWHM of H$\beta$ cites from \cite{Peterson2014} (after a private communication with B. M. Peterson). \\
``$b$" The line dispersions of these broad emission lines from the rms spectra were calculated from the virial products 
(see Table 6 of \citealt{Peterson2014}) by combining the time lags ($\tau$), 
because we do not directly find these values from the published papers. \\
References. (1) This work, (2) \cite{Peterson1998}, (3) \cite{Collin2006}, (4) \cite{Bentz2013}, (5) \cite{Denney2010}, (6) \cite{Peterson2014}. 
}
\end{deluxetable*}

\begin{figure*}[ht!]
\centering
\includegraphics[angle=0,width=0.8\textwidth]{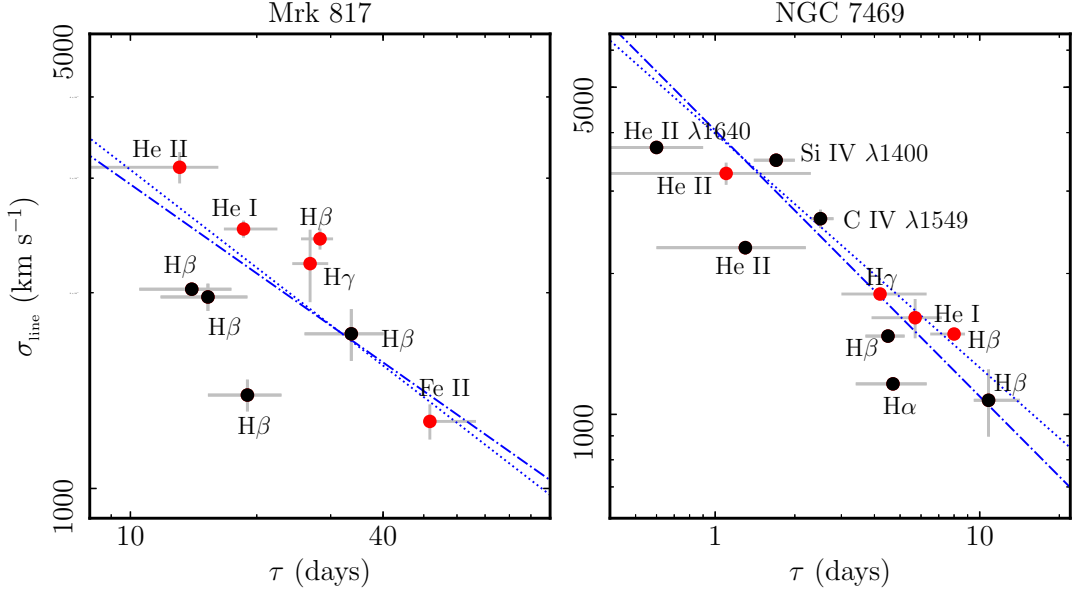}
\caption{\footnotesize 
Relationship between broad emission-line widths and their time lags for Mrk~817 and NGC~7469. 
The red circles are new measurements from the present campaign, 
the black circles are previous measurements (references refer to Tabel~\ref{tab_sum}). 
Emission-line names are noted near its position, 
where the names of UV broad emission lines have the information of wavelength (i.e., He~{\sc ii}~$\lambda1640$, C~{\sc iv}~$\lambda1549$ and Si~{\sc iv}~$\lambda1400$), 
and the names of optical broad emission lines without the information of wavelength (i.e., H$\gamma$, H$\beta$, He~{\sc ii}, He~{\sc i} and Fe~{\sc ii} multiplets).
The dot-dashed lines are the fit to the relationship ${\rm log}\sigma_{\rm line}=a+b~{\rm log}\tau$, 
which have slopes $\rm b=-0.45\pm0.44$ for Mrk~817 and $\rm b=-0.55\pm0.15$ for NGC~7469. 
The dotted lines are the best fit with a forced virial slope of $b=-0.5$.
}
\label{fig_vir}
\end{figure*}

\subsection{Line Profile Measurements and Virial Relationship} 
\label{sec_vir}
The line width of the broad emission line can be characterized by either the full width at half maximum (FWHM) or the line dispersion ($\sigma_{\rm line}$). 
In order to determine the best value of the line width and its uncertainty from net broad emission lines (including H$\gamma$, H$\beta$, He~{\sc ii} and He~{\sc i}) 
which were constructed in the spectral fitting (Section~\ref{sec_lc}), 
we use Monte Carlo simulations similar to those used when determining the time lag from the CCF 
to generate 200 mean and rms spectra from 200 randomly chosen subsets of the spectra. 
From which we measure FWHM and $\sigma_{\rm line}$, and obtain the distributions of FWHM and $\sigma_{\rm line}$. 
We adopt the mean values of FWHM and $\sigma_{\rm line}$ from the distributions and adopt their standard deviations as uncertainties. 
We compare the line widths of O~{\sc iii}~$\lambda$5007 (FWHM$=$360~km~$\rm s^{-1}$ for NGC~7469, and FWHM$=$330~km~$\rm s^{-1}$ for Mrk~817) 
measured from the high-resolution spectrum (see \citealt{Whittle1992}) with those from our spectra, 
calculating the instrumental broadening. 
Including the line dispersion of Fe~{\sc ii} for Mrk~817 calculated from the best-fit parameter, 
we list the broadening-corrected line widths in Table~\ref{tab_sum} for the present campaign. 

In order to investigate the virial relationship for individual target, 
we compile line widths and time lags of the broad emission lines (including optical and UV band) measured by previous RM campaigns (see Table~\ref{tab_sum}). 
All time lags in Col.~8 were obtained from the centroid of CCF. 
Because all campaigns almost have the measurements of line dispersion ($\sigma_{\rm line}$, in Col.~7) from the rms spectrum, 
we use the line dispersions and in combination with the time lags to examine the virial relationship of the BLR. 
We include the measurement of Fe~{\sc ii} for Mrk~817 in this examination, where $\sigma_{\rm line}$ of Fe~{\sc ii} is the mean value of the best-fit parameters. 
The left and right panels of Figure~\ref{fig_vir} show the relationship between line widths and time lags for Mrk~817 and NGC~7469, respectively. 
The dot-dashed lines are the fit to the relationship ${\rm log}\sigma_{\rm line}=a+b~{\rm log}\tau$, 
they give slopes $\rm b=-0.45\pm0.44$ for Mrk~817 and $\rm b=-0.55\pm0.15$ for NGC~7469. 
The dotted lines are the best fit with a forced virial slope of $b=-0.5$. 
In fact, these examination results for both targets are consistent with the virial prediction of the BLR. 

\subsection{Black Hole Mass and Accretion Rates} 
\label{sec_bh}
Under assumption that the BLR are dominated by the gravity of the central SMBH, 
we first calculate the virial product of Mrk~817 and NGC~7469 by 
\begin{equation}
\Hat{M}_{\bullet}=\frac{c\tau_{_{\rm H\beta}} V^2}{G}, 
\label{eq_vp}
\end{equation}
where $c\tau_{_{\rm H\beta}}$ is the size of the BLR, $c$ is the speed of light, $G$ is the gravitational constant, 
the BLR velocity or line width $V$ is either full width at half maximum (FWHM) or line dispersion ($\sigma_{\rm line}$) of 
the broad emission lines in the mean or rms spectrum. 
As other campaigns, we use the RM measurements of H$\beta$ to estimate the virial product that is actually responding to the SMBH mass. 
The line-width measurement from the rms spectrum eliminates contamination from constant narrow line. Meanwhile, 
\cite{Peterson2011} gave some evidences that line dispersion of $\sigma_{\rm line}$ produces less biased mass measurements than FWHM. 
Therefore, we calculate the virial product of Mrk~817 and NGC~7469 using the line dispersion of $\sigma_{\rm line}$ from the rms spectrum. 
We list $\Hat{M}_{\bullet}$ of the present campaign in Table~\ref{tab_rm3}. 

Totally, Mrk 817 and NGC 7469 have five and three H$\beta$-based RM measurements (see Table~\ref{tab_sum}). 
For comparison, we also calculate the virial products of the previous H$\beta$-based campaigns for both targets 
though the same manner with the present campaign and tabulate them into Table~\ref{tab_rm3}. 
For completeness, Table~\ref{tab_rm3} lists the H$\beta$ time lag ($\tau_{\rm H\beta}$) and its line width ($\sigma_{\rm line}$ from the rms spectrum), 
both parameters are used to calculate the virial product. 
For Mrk~817, the virial product obtained in the present campaign is consistent with the 1995's campaign (observation epoch is 03/02/1995$\sim$24/07/1995) within the uncertainties. 
For NGC~7469, the virial product obtained in the present campaign is same with \cite{Peterson2014}'s campaign. 

At the present time, the geometric structure,  kinematics and origins of the BLR are still unknown, 
the dimensionless virial factor $f_{\rm BLR}$ is introduced to attenuate effects of these unknown information on the measurement of SMBH mass ($M_{\bullet}$). 
\cite{Graham2011} evaluated the factor $f_{\rm BLR}$ taking into account the morphology of the host galaxies the first, 
and found that the $f_{\rm BLR}$ for barred galaxies is three times lower than that for non-barred galaxies. 
\citet{Ho2014} spent a lot of effort to reevaluate the $f_{\rm BLR}$ for the RM AGN sample taking into account the bulge types 
of host galaxies, and found that the systematic difference in $f_{\rm BLR}$ between 
barred and non-barred galaxies qualitatively resembles the dependence on bulge type. 
Who suggested that pseudo-bulge has a lower $f_{\rm BLR}$ than in classical bulge. 
For $\sigma_{\rm line}$ measured from the rms spectrum, 
$f_{\rm BLR}=3.2\pm0.7$ for pseudo-bulges, whereas $f_{\rm BLR}=6.3\pm1.5$ for classical bulges. 
For FWHM measured from the rms spectrum, 
$f_{\rm BLR}=0.7\pm0.2$ for pseudo-bulges, whereas $f_{\rm BLR}=1.5\pm0.4$ for classical bulges. 
More values of $f_{\rm BLR}$ have been suggested, we will give extra discussion in Section~\ref{sec_dis2}. 

In view of the bulge types of host galaxies of Mrk~817 and NGC~7469 are known (pseudo-bulge; \citealt{Ho2014}), 
and line dispersion of $\sigma_{\rm line}$ produces less biased mass measurements than FWHM \citep{Peterson2011}, 
we use $f_{\rm BLR}=3.2$ and the $\sigma_{\rm line}$ from the rms spectrum to calculate the SMBH mass for all RM campaigns via 
\begin{equation}
M_{\bullet}=f_{\rm BLR}\times\Hat{M}_{\bullet}, 
\label{eq_mbh}
\end{equation}
and list the results in Table~\ref{tab_rm3}. 
The uncertainties of SMBH masses only account for the errors of time lag and line width of the H$\beta$. 
It should be mentioned that, based on Atacama Large Millimeter$/$submillimeter Array (ALMA) observations, 
the dynamical method of estimating SMBH mass from the centre of AGN was developed in recent years. 
For example, based on the dynamical modelling of the atomic-[CI](1-0) data, \cite{Nguyen2021} suggested that 
the SMBH mass in NGC~7469 is $1.78\times10^{7}M_{\odot}$. 
Which is almost consistent with the RM-based SMBH mass of NGC~7469 from the present campaign and \cite{Peterson2014}'s campaign (see Col.~8 of Table~\ref{tab_rm3}). 

In light of the standard model of accretion disc \citep{Shakura1973}, 
the dimensionless accretion rates are related to the optical luminosity at 5100~\AA~and SMBH mass via \citep{Du2015} 
\begin{equation} 
\mathdotM=\frac{\dot{M}_{\bullet}}{L_{\rm Edd}c^{-2}}=20.1\left(\frac{\ell_{44}}{\cos i}\right)^{3/2}M_7^{-2}, 
\label{eq_mdot}
\end{equation}
\\
where $L_{\rm Edd}$ is the Eddington luminosity, $c$ is the speed of light, $\dot{M}_{\bullet}$ is the mass accretion rates, 
$i$ is the inclination of the accretion disc. We take $\cos i=0.75$, which represents a mean disc inclination for a type 1 AGN, 
$\ell_{44}=L_{5100}/10^{44}~\ergs$ is optical luminosity at 5100~\AA, $M_7=\bhm/10^7\sunm$ is SMBH mass. 
In Section~\ref{sec_lc}, we obtained AGN continuum light curves at 5100~\AA~of Mrk~817 and NGC~7469 for the present campaign, 
which have eliminated the contamination of host galaxy by spectral fitting and decomposition. 
We calculate the averaged flux ($F_{\rm 5100}$) and standard deviation (as the uncertainty of $F_{\rm 5100}$) from their light curves, and list in Table~\ref{tab_rm3}. 
For the previous campaigns of both targets, we compile AGN continuum flux at 5100~\AA~($F_{\rm 5100}$) and tabulate them into Table~\ref{tab_rm3}, 
the contaminations of host galaxy on these AGN continuum flux at 5100~\AA~have been eliminated by decomposing {\it HST} image (see \citealt{Bentz2009,Bentz2013,Peterson2014}). 
For both targets, we calculate the optical luminosity at 5100~\AA~($L_{\rm 5100}$) using AGN continuum flux at 5100~\AA, and list in Table~\ref{tab_rm3}. 
Using optical luminosity and SMBH mass (Col. 8 of Table~\ref{tab_rm3}), we calculate the dimensionless accretion rates for all RM campaigns and list in Table~\ref{tab_rm3}. 

Following \cite{Du2015}, we classify accretion of AGN into sub-Eddington and super-Eddington regimes by $\mathdotM=3$, 
beyond which the radial advection of accrete flows is not negligible, the inner parts of disc break the standard model of accretion disc \citep{Laor1989}. 
The AGN with high accretion rates ($\mathdotM\gtrsim 3$) are thought to be powered by slim disk \citep{Abramowicz1988}. 
We find that Mrk~817 accretes in sub-Eddington in almost all monitoring periods ($\dot{M}_{\bullet}<3 \times L_{\rm Edd}c^{-2}$), 
while NGC~7469 accretes in super-Eddington ($\dot{M}_{\bullet}>3 \times L_{\rm Edd}c^{-2}$). 
We will further discuss the measurements about the SMBH mass and accretion rates in Section~\ref{sec_dis2}. 

\begin{deluxetable*}{lccccccccc}
  \tablecolumns{10}
  \tabletypesize{\scriptsize}
  \setlength{\tabcolsep}{3pt}
  \tablewidth{4pt}
  \tablecaption{Summary of H$\beta$ reverberation mapping for Mrk~817 and NGC~7469\label{tab_rm3}}
  \tablehead{
  \colhead{Object}&
  \colhead{Observation epoch} &
  \colhead{$\tau_{_{\rm H\beta}}$} &
  \colhead{$\sigma_{\rm line}$}&
  \colhead{$F_{\rm 5100}$} &
  \colhead{$L_{\rm 5100}$} &
  \colhead{$\Hat{M}_{\bullet}$} &
  \colhead{$M_{\bullet}$} &
  \colhead{$\dot{\mathscr{M}}$} &
  \colhead{References} \\
  \colhead{} &
  \colhead{(dd/mm/yy$-$dd/mm/yy)} &
  \colhead{(days)} &
  \colhead{~(km~s$^{-1}$)} &
  \colhead{($\times10^{-15}$)} &
  \colhead{($\times10^{43}\rm ~erg~s^{-1}$)}  &
  \colhead{($\times10^{7} M_{\odot}$)} &
  \colhead{($\times10^{7} M_{\odot}$)} &
  \colhead{} &
  \colhead{} \\
  \colhead{(1)} &
  \colhead{(2)} &
  \colhead{(3)} &
  \colhead{(4)} &
  \colhead{(5)} &
  \colhead{(6)} &
  \colhead{(7)} &
  \colhead{(8)} &
  \colhead{(9)} &
  \colhead{(10)} 
  }
\startdata
Mrk~817     &20/12/2018$-$20/05/2019&$28.3^{+2.1}_{-1.8}$ &$2143\pm56$&  $3.77\pm0.30$ &$4.82\pm0.38$&$2.54^{+0.23}_{-0.21}$ &$8.12^{+0.73}_{-0.67}$ & 0.16   & 1 \\   
                   &25/03/2007$-$19/07/2007&$14.0^{+3.4}_{-3.5}$ &$2025\pm5$&   $5.17\pm0.14$ &$6.62\pm0.18$&$1.12^{+0.27}_{-0.28}$ &$3.59^{+0.87}_{-0.90}$ & 1.29   & 2\\  
                   &03/02/1995$-$24/07/1995&$33.6^{+6.5}_{-7.6}$ &$1729\pm158$&$3.52\pm0.15$ &$4.50\pm0.19$&$1.96^{+0.52}_{-0.57}$ &$6.28^{+1.67}_{-1.83}$ & 0.24   & 2\\  
                   &20/02/1994$-$23/06/1994&$15.3^{+3.7}_{-3.5}$ &$1970\pm95$&  $3.51\pm0.15$ &$4.49\pm0.19$&$1.16^{+0.30}_{-0.29}$ &$3.71^{+0.97}_{-0.92}$ & 0.68   & 2\\  
		  &12/01/1993$-$11/08/1993&$19.0^{+3.9}_{-3.7}$ &$1392\pm78$&  $4.61\pm0.20$ &$5.90\pm0.26$&$0.72^{+0.17}_{-0.16}$ &$2.30^{+0.53}_{-0.52}$ & 2.65   & 2\\   
NGC~7469 &12/10/2019$-$31/12/2019&$8.0^{+0.8}_{-1.5}$  &$1485\pm34$&  $12.40\pm1.50$&$4.12\pm0.50$&$0.35^{+0.04}_{-0.06}$ &$1.10^{+0.12}_{-0.21}$ & 6.72& 1 \\   
		  &31/08/2010$-$28/12/2010&$10.8^{+3.4}_{-1.3}$&$1274\pm126$& $7.84\pm0.60$ &$2.60\pm0.20$&$0.34^{+0.13}_{-0.08}$ &$1.09^{+0.41}_{-0.25}$ & 3.42& 3   \\           
		  &02/06/1996$-$30/07/1996&$4.5^{+0.7}_{-0.8}$  &$1456\pm207$&  $5.14\pm0.50$ &$1.71\pm0.17$&$0.19^{+0.06}_{-0.06}$ &$0.60^{+0.19}_{-0.20}$ & 6.14& 4           
\enddata   
\tablecomments{\footnotesize
The $\sigma_{\rm line}$ is measured from the rms spectrum. 
In order to make it easy for readers to calculate the optical luminosity with the cosmological parameters they are used to taking, 
we also list AGN continuum flux at 5100~\AA~($F_{\rm 5100}$) of each campaign in units of~erg~s$^{-1}$~cm$^{-2}$~\AA$^{-1}$, 
which has eliminated the contamination of starlight from host galaxy. 
In fact, $F_{\rm 5100}$ in this table is the averaged continuum flux at 5100~\AA~over each RM monitoring period. 
$L_{\rm 5100}$ in Col. (5) is the optical luminosity at 5100~\AA. 
We use H$\beta$ time lag (Col. 3) and $\sigma_{\rm line}$ (Col. 4) to calculate the virial product (Col. 7). 
The bulge type of host galaxies for Mrk~817 and NGC~7469 is pseudo-bulge \citep{Ho2014}, 
using $f_{\rm BLR}=3.2$ and in combination with virial product yields SMBH mass (Col. 8). 
We use optical luminosity (Col. 6) and SMBH mass to estimate the accretion rates (Col. 9). 
\\
References. (1) This work, (2) \cite{Peterson1998}, (3) \cite{Peterson2014}, (4) \cite{Collier1998}.} 
\end{deluxetable*}

 \begin{figure}[ht!]
\centering
\includegraphics[angle=0,width=0.495\textwidth]{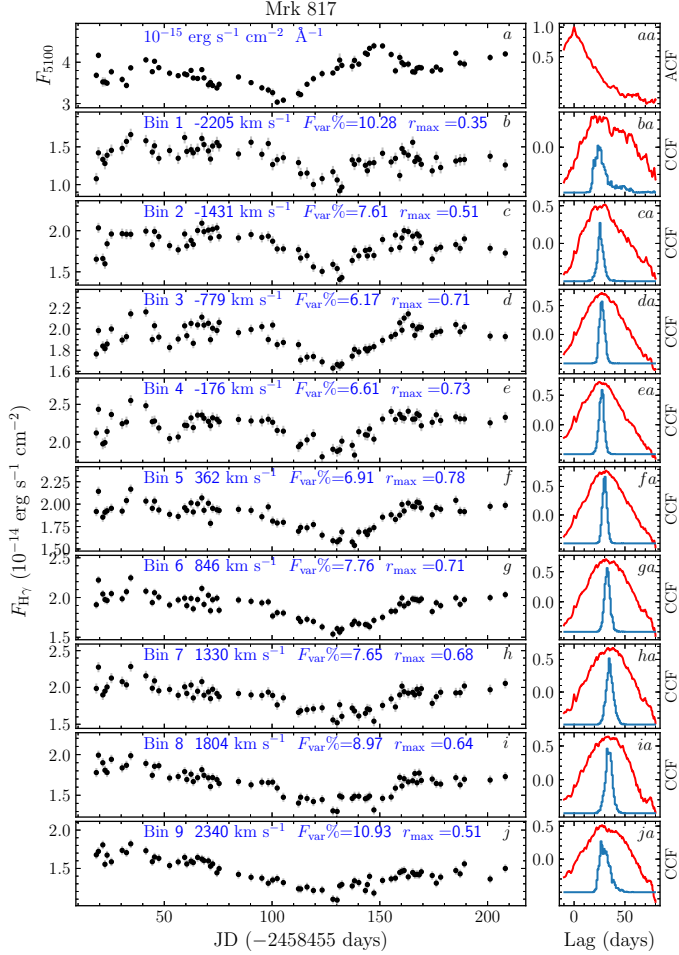}
\caption{\footnotesize
Velocity-resolved reverberation mapping of Mrk~817. 
The left panels (a)$-$(j) show the light curves of AGN continuum at 5100~\AA~
and the velocity-resolved light curves of broad $\rm H\gamma$ emission line, respectively. 
The right panels (aa)$-$(ja) correspond to the ACF of AGN continuum at 5100~\AA~and the 
CCF between the light curves of each velocity bin (b)$-$(j) and the continuum variation (a), respectively. 
We note the central velocity of each bin ({\tt Bin~1$-$9}; from blue side to red side of the broad H$\gamma$ line), 
variability amplitude ($F_{\rm var}\%$) of the light curves, the maximum correlation coefficients ($r_{\rm max}$) in panels (b)$-$(j). 
In panels (ba)$-$(ja), we plot the ICCF in red, and the cross-correlation centroid distribution (i.e., CCCD) in blue.
}
\label{fig_vlp1}
\end{figure}

\begin{figure*}[ht!]
\centering
\includegraphics[angle=0,width=0.495\textwidth]{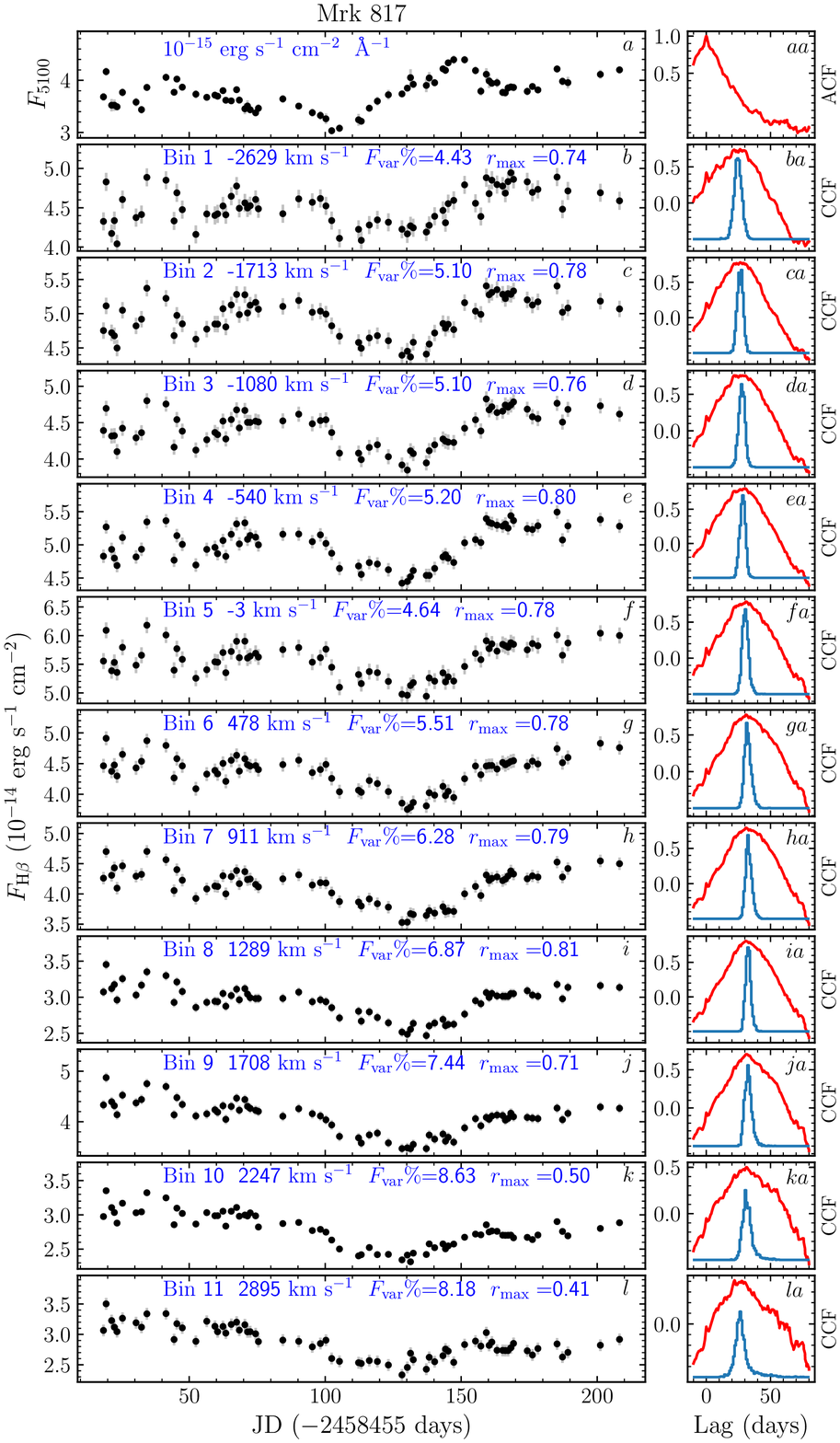}
\includegraphics[angle=0,width=0.495\textwidth]{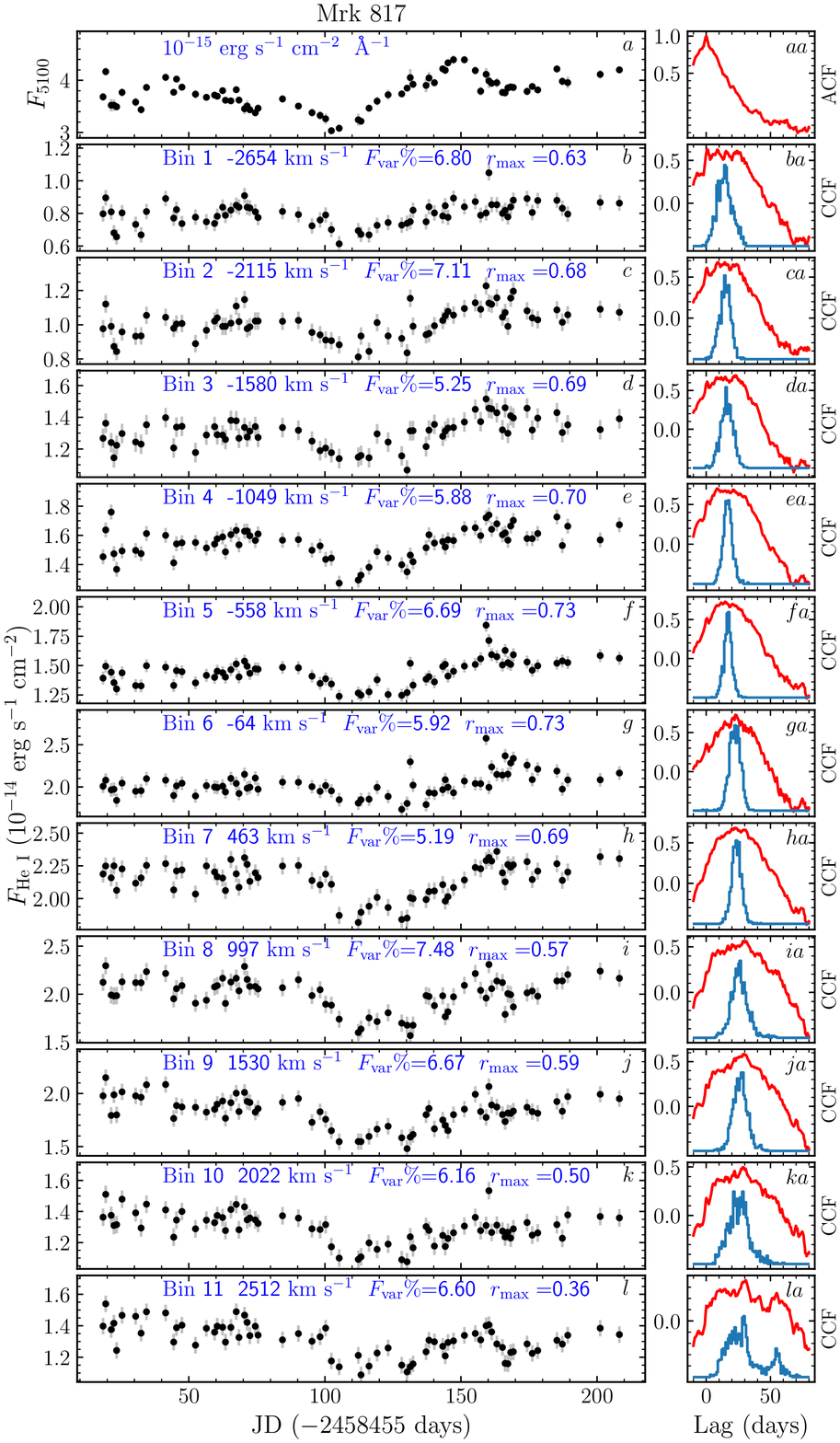}
\caption{\footnotesize
Same as Figure~\ref{fig_vlp1}, but for the broad H$\beta$ and He~{\sc i} emission lines of Mrk~817. 
}
\label{fig_vlp2}
\end{figure*}

\begin{figure*}[ht!]
\centering
\includegraphics[angle=0,width=0.495\textwidth]{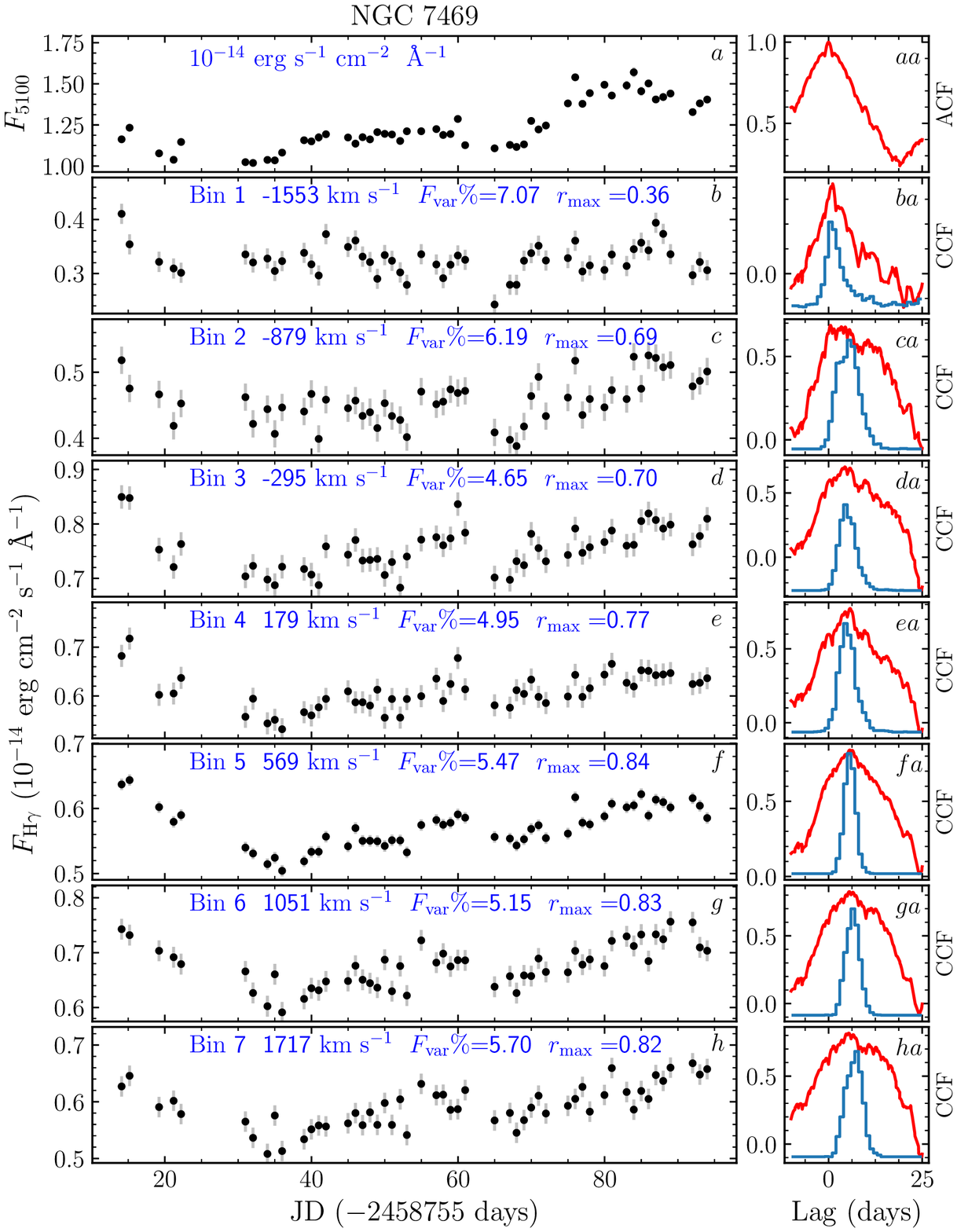}
\includegraphics[angle=0,width=0.495\textwidth]{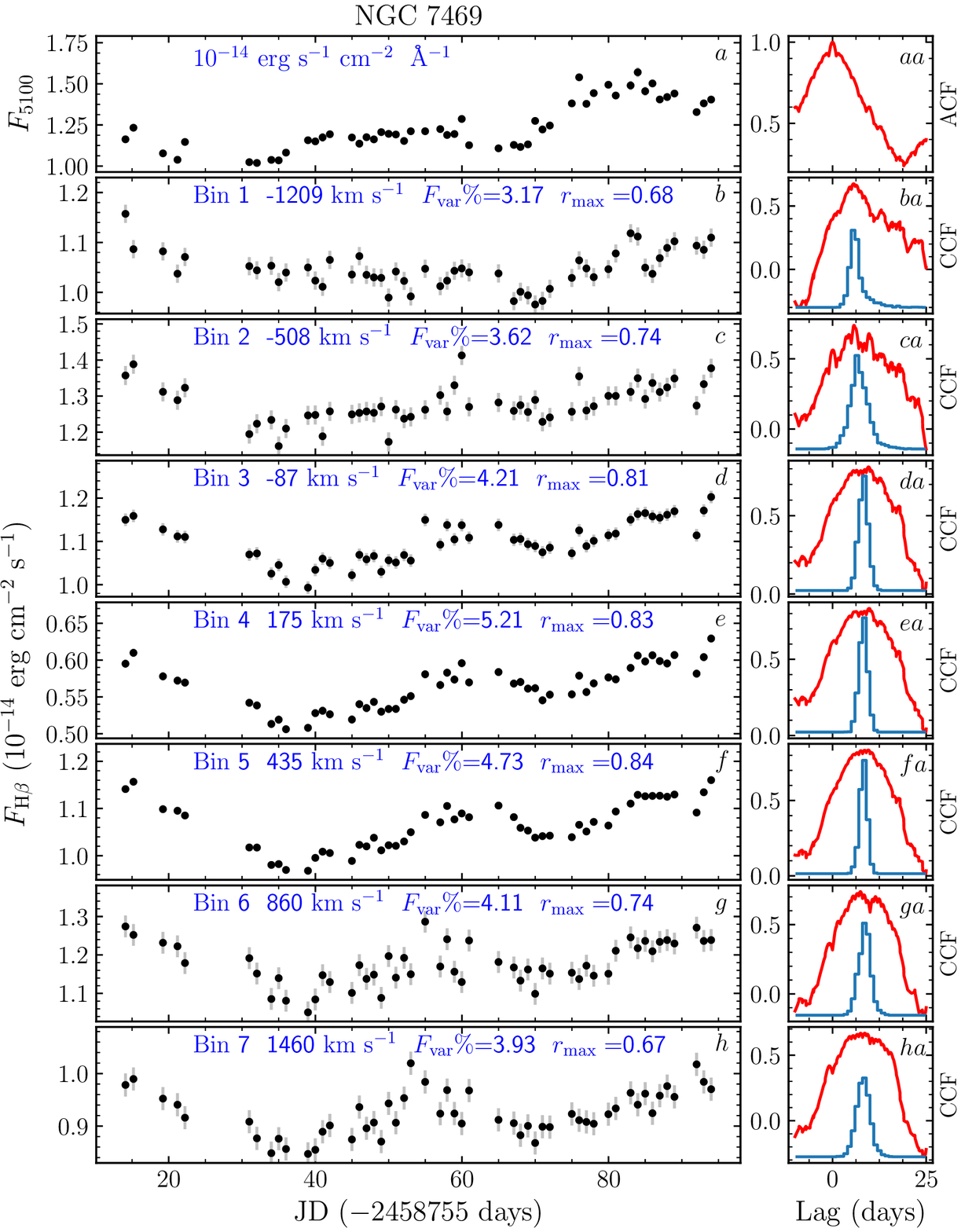}
\caption{\footnotesize
Same as Figure~\ref{fig_vlp1}, but for the broad H$\gamma$ and H$\beta$ emission lines of NGC~7469. 
}
\label{fig_vlp3}
\end{figure*}

\begin{figure*}[ht!]
\centering
\includegraphics[angle=0,width=0.32\textwidth]{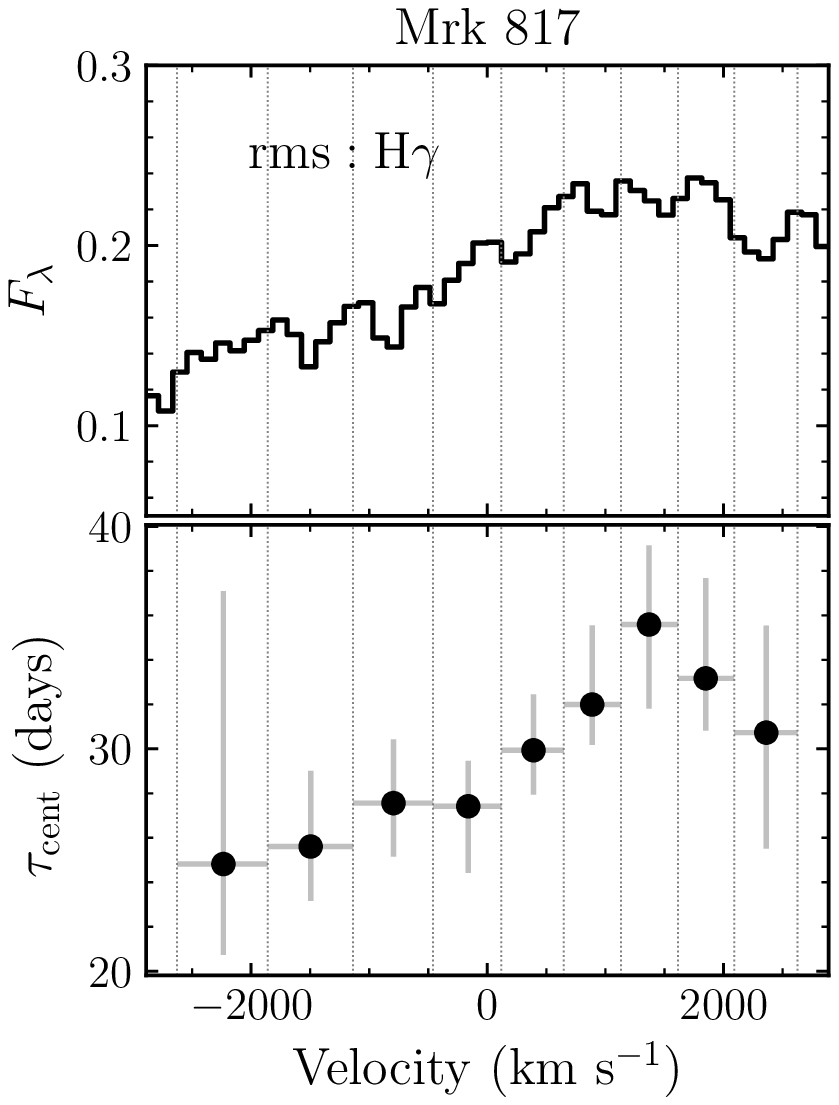}
\includegraphics[angle=0,width=0.32\textwidth]{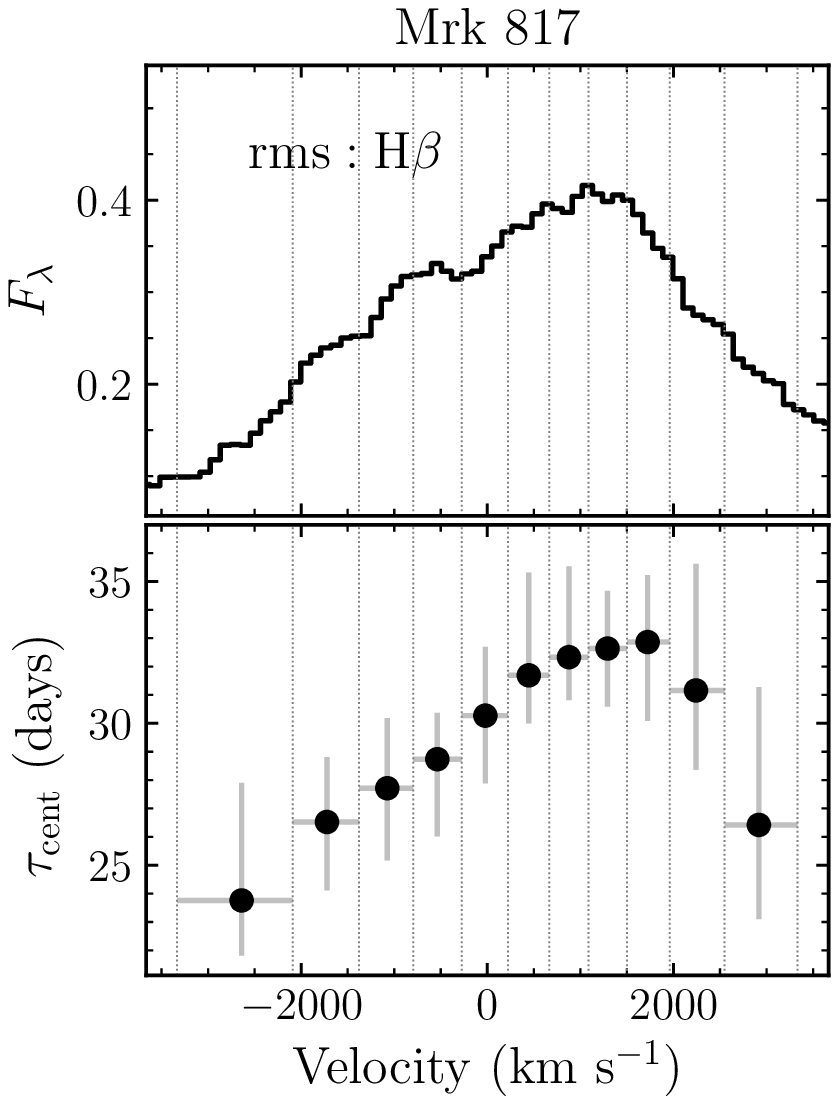}
\includegraphics[angle=0,width=0.32\textwidth]{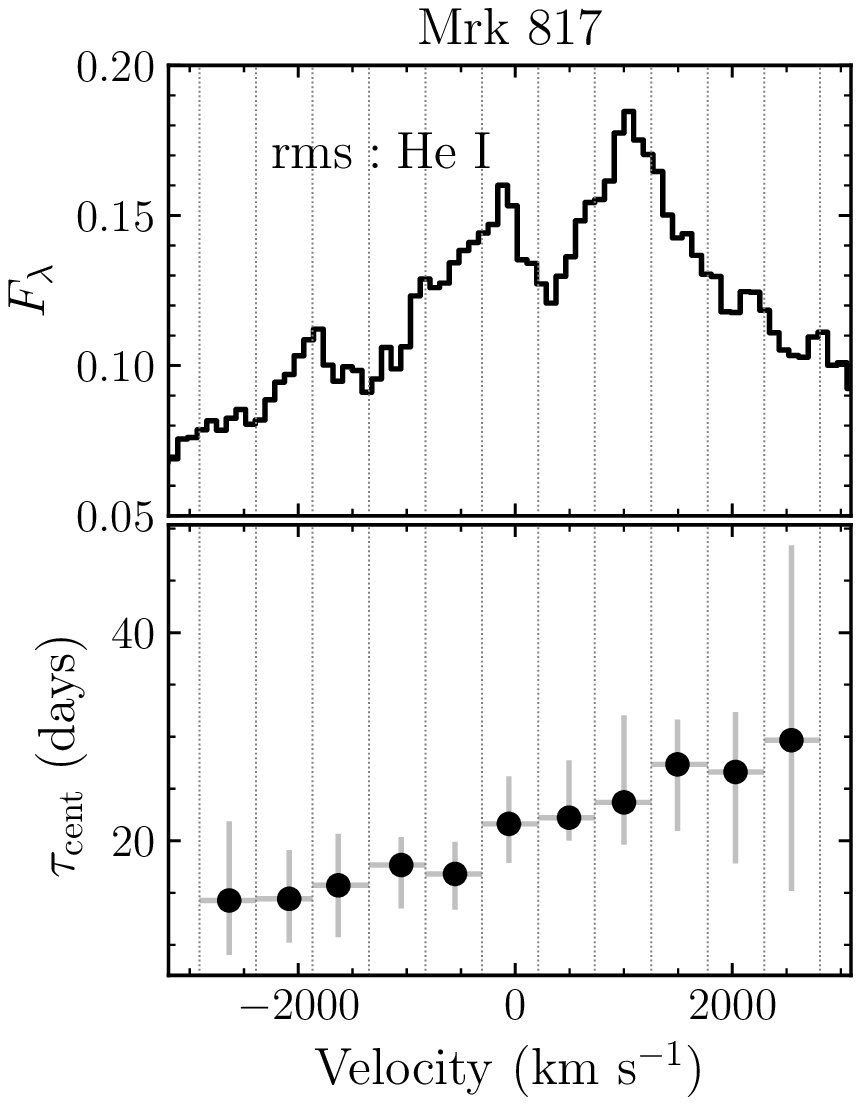}\\
\includegraphics[angle=0,width=0.33\textwidth]{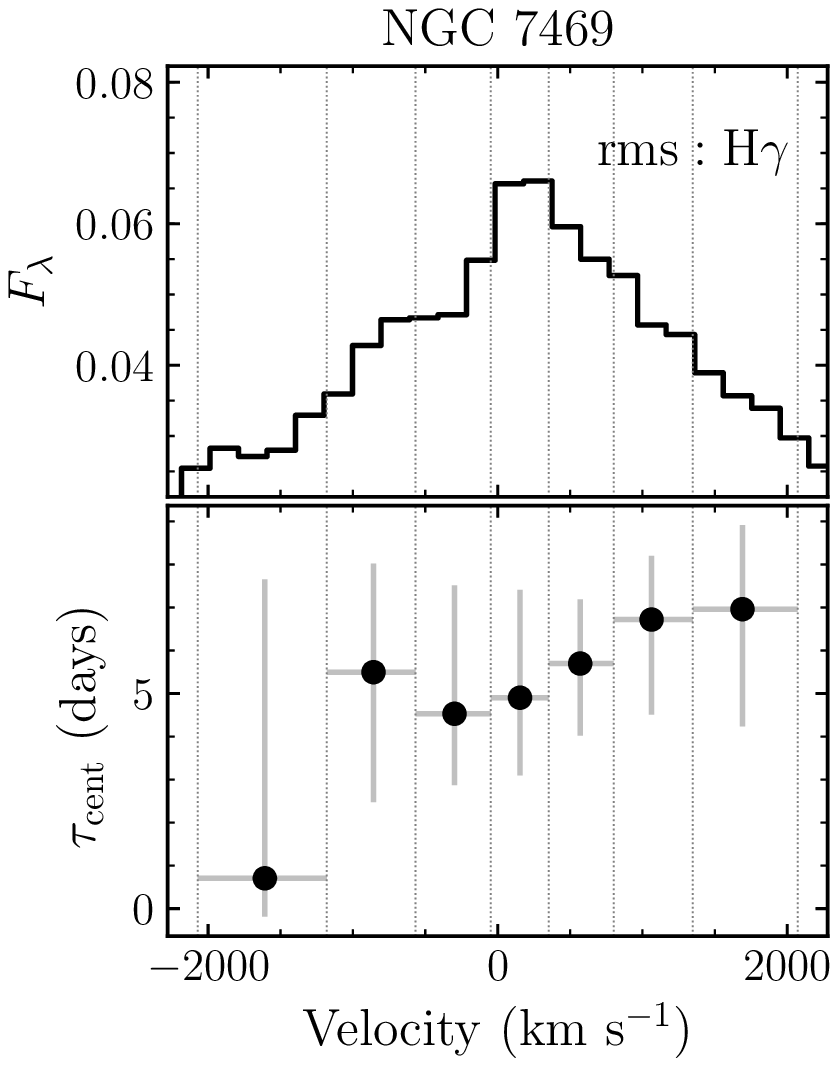}
\includegraphics[angle=0,width=0.34\textwidth]{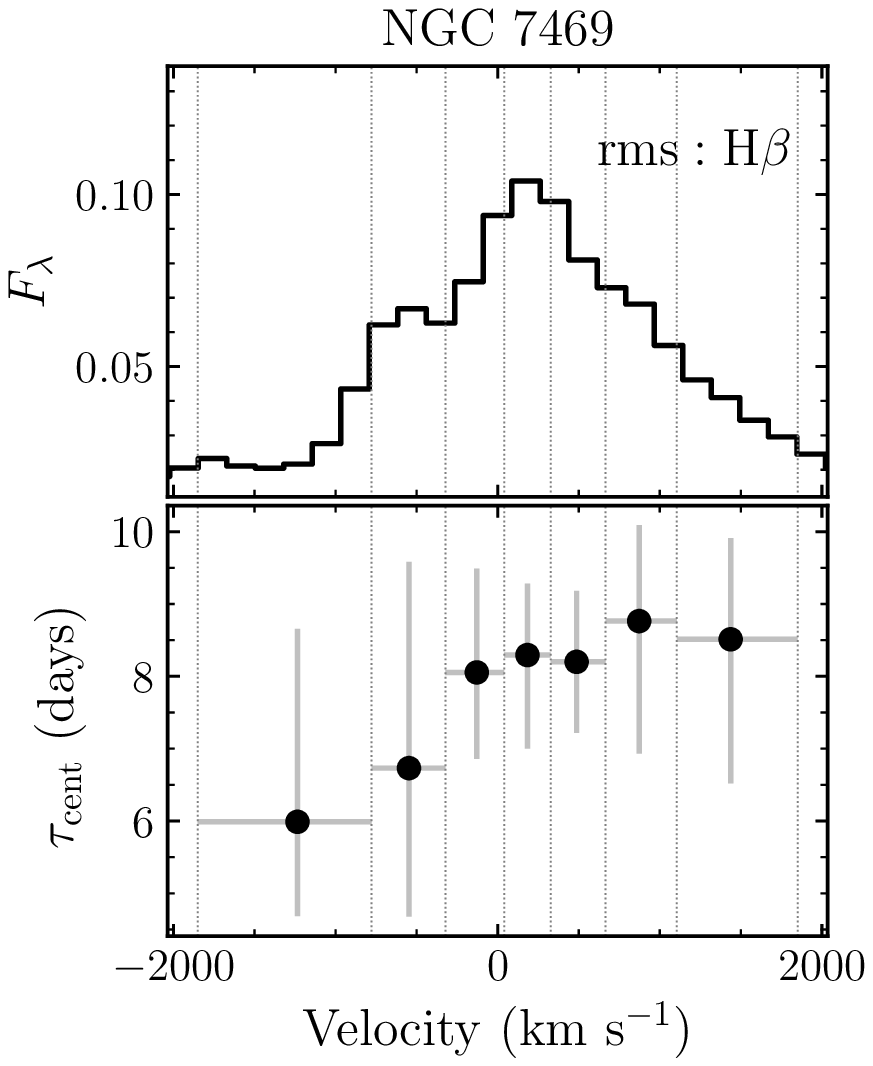}
\caption{\footnotesize
The results of velocity-resolved reverberation mapping. In each case, top panel is the rms profiles of broad emission lines 
(H$\gamma$, H$\beta$, He~{\sc i} for Mrk~817; H$\gamma$ and H$\beta$ for NGC~7469), 
bottom panel is the velocity-resolved lag profiles. 
}
\label{fig_vlp4}
\end{figure*}

\subsection{Velocity-resolved Reverberation Mapping}
\label{sec_vm}
The time lags measured between the varying AGN continuum and the total flux variations of the broad emission lines 
just give out the radii averaged by the emissivity function of the BLR. The velocity of the BLR gas should be a function of radius, 
the kinematic discrepancies of the different BLR gas that respond to the continuum variations should be presented in the velocity-resolved time lags. 
Therefore, velocity-resolved lag profiles were widely used to expose the kinematic structures of the BLR in many AGNs. 
In the previous four observing seasons of Mrk~817, only \cite{Denney2010} constructed the velocity-resolved lag profiles of H$\beta$, 
which marginally show an outflow signature since the time lags of first four velocity bins are negative (see Figure 5 of \citealt{Denney2010}). 
In order to meet the goal of recovering velocity-delay maps from the data, \cite{Peterson2014} developed an intensively spectroscopic monitoring campaign of NGC~7469 in 2010, 
but who didn't achieve this goal because of a very low level of variability during the monitoring periods. 

To recover velocity-resolved lag profiles from the data of Mrk~817 and NGC~7469, 
we first obtain the net broad emission lines by eliminating other fitted components, 
and calculate the rms spectrum (i.e., variable spectrum) of broad emission lines using Equation~\ref{eq_rsp}. 
It should be noted that, in the past velocity-resolved reverberation mapping, 
the most widely used method of binning spectrum is to divide the variable spectrum into bins with equal flux, 
or equal velocity width sometimes. 
Following widely used binning method of the rms spectrum (e.g., \citealt{Denney2009,Grier2013a}), 
we then separately divide emission-line flux into several velocity-space bins with equal flux 
in the rms spectrum, finally obtain the light curves of every bins by integrating the continuum-subtracted flux in the bin. 
We plot the corresponding light curves in the left panels of Figures~(\ref{fig_vlp1},~\ref{fig_vlp2},~\ref{fig_vlp3}), 
where Figures~\ref{fig_vlp1}~and~\ref{fig_vlp2} are the velocity-resolved light curves of H$\gamma$, H$\beta$ and He~{\sc i} of Mrk~817, 
and Figure~\ref{fig_vlp3} is the velocity-resolved light curves of H$\gamma$ and H$\beta$ of NGC~7469. 
These light curves are numbered with {\tt Bin} number (i.e., Bin 1 to $N$ from blue wing, 
line core to red wing of the broad emission lines) along with the central velocity of each bin. 
The time lags of every velocity-resolved light curves with respect to the varying AGN continuum at 5100~\AA~and 
associated uncertainties are determined using same procedures as described in Section~\ref{sec_lag}. 

The results of cross-correlation analysis are plotted in the right panels of Figures~(\ref{fig_vlp1},~\ref{fig_vlp2},~\ref{fig_vlp3}). 
We calculate variability amplitudes $F_{\rm var}$ of the velocity-resolved light curves using Equation~(\ref{eq_fvar}), 
and note these values in the left panels of Figures~(\ref{fig_vlp1},~\ref{fig_vlp2},~\ref{fig_vlp3}) along with the maximum cross-correlation coefficient. 
The velocity-resolved lag profiles (VLPs) for both targets are presented in Figure~\ref{fig_vlp4}. For each case, the top panel shows the rms spectrum of broad emission line, 
the bottom panel simply shows the velocity-resolved time lags as a function of velocity. 
For Mrk~817, VLPs including H$\gamma$, H$\beta$ and He~{\sc i} have similar structure that the time lags of blue wing are smaller than red wing. 
This trend is similar with the VLPs of broad H$\beta$ line recovered by \cite{Denney2010}. 
For NGC~7469, VLPs including H$\gamma$ and H$\beta$ also have similar structure that the time lags of blue wing are also smaller than red wing. 
We also try to divide the rms spectrum into bins with equal velocity width, 
find the VLPs for both targets are consistent with Figure~\ref{fig_vlp4}, so we only present above results. 
According the simulated velocity-resolved maps (e.g., \citealt{Bentz2009,Bentz2010a,Grier2013a}), in some degree, 
these signatures demonstrate that the BLR in both targets could has some outflowing gas during the monitoring periods. 
Further discussions for the kinematics of BLR are presented in Section~\ref{sec_dis1}. 
The investigation of two-dimensional velocity-delay maps (e.g., \citealt{Horne2004}) and 
dynamical modelling of the BLR (e.g., \citealt{Pancoast2014,Li2018}) remain the focus a separate work. 

\section{Discussion}
\label{sec_dis}
\subsection{Kinematics of the BLR}
\label{sec_dis1}
In Section~\ref{sec_vm}, we find that Mrk~817 and NGC~7469 have very similar velocity-resolved lag profiles that 
the time lags in the red wing are slightly larger than the time lags in the blue wing, 
which is consistent with the kinematic signature of outflow gas motion according to the simulated velocity-resolved maps 
(\citealt{Bentz2009,Bentz2010a,Grier2013a}). 
However, during the spectral fitting, 
we find that the broad H$\beta$ emission lines of Mrk~817 and NGC~7469 are red asymmetric (also refer to \citealt{Shapovalova2017} for NGC~7469), 
that is the broad H$\beta$ emission lines show the red-shifted signature for both targets. 
The blue shift of broad emission line with respect to the [O~{\sc iii}]~$\lambda5007$ is explained as the existence of outflowing emitters in BLR \citep{Richards2011}, 
while the red shift of broad emission line with respect to the [O~{\sc iii}]~$\lambda5007$ could originate from inflowing emitters (\citealt{Hu2008,Shapovalova2017}). 
If the latter holds, 
the emitters of broad H$\beta$ emission line in Mrk~817 and NGC~7469 seem to have inflowing components during the monitoring period. 
This is ambivalent with above kinematic signatures. 
Certainly, as many observational reasons and theoretical model suggestion, 
the unresolved BLR could has a more complex structures than suggestions of the simulated velocity-resolved maps, 
such as the BLR could be a multi-structure combination (e.g, \citealt{Murray1995,Denney2010,Grier2013a,Zhangsh2019}). 

In addition, about half of Seyfert galaxies exit multi-phased winds 
(\citealt{Reynolds1995,Krolik2001,Leighly2004,Tombesi2013,Mehdipour2018,Williamson2020,Matthews2020}), 
the complex outflowing winds were also observed in Mrk~817 and NGC~7469 from the X-ray and ultraviolet spectrum. 
These variable winds (such as filling factors and column density) 
may serve to scatter or shield the interactions between the BLR and the central engine. 
Simultaneous and multi-band spectroscopic monitoring campaign, such as the upcoming AGN STORM 2 campaign (\citealt{Kara2021arXiv}), 
should give some limits to the potential connection between the multi-phased winds and the BLR. 

\subsection{SMBH mass and accretion rates}
\label{sec_dis2}
The sphere of influence (SOI) for the case of SMBH inhabiting AGN 
is defined as the region of space within which the stars and gases (or clouds) are dominated by the gravitational potential of the SMBH \citep{Ferrarese2005}. 
Its radius is given by $r_{\rm SOI} \sim G M_{\bullet} / \sigma_{*}^{2} \sim 11.2~(M_{\bullet}/10^{8}M_{\odot}) / (\sigma_{*}/ {\rm km~s^{-1}})^{2}~{\rm pc}$, 
where $\sigma_{*}$ is the velocity dispersion of the surrounding stars, and other symbols are same with above definition. 
The stellar velocity dispersion of Mrk~817's and NGC~7469's bulge are 120$\pm$15~km~s$^{-1}$ and 131$\pm$5~km~s$^{-1}$ \citep{Ho2014}, respectively. 
We find that this radius ($r_{\rm SOI}$) approximates to 20~pc for Mrk~817 and 1.0~pc for NGC~7469, which is far greater than the BLR size 
($\sim$30 light-days for Mrk~817, and $\sim$10 light-days for NGC~7469). 
Although gas, unlike stars, can easily be accelerated by non gravitational forces, 
the BLR gas or cloud inevitably locates in the inner region of SOI 
and consequentially dominates by the gravity of the central SMBH for Mrk~817 and NGC~7469(see Section~\ref{sec_vir} and Figure~\ref{fig_vir}). 
At the same time, the origin of unresolved BLR (e.g., disc, infall, outflow or wind) is a controversial subject that is not yet resolved 
(e.g., \citealt{Collin-Souffrin1987,Murray1995,Wang2017,Czerny2017,Adhikari2018,Baskin2018}). 
Therefore, to within the astrometric errors, 
we believe that the SMBH mass should be restricted reasonably by the reverberation-mapping measurements 
even given above kinematic signatures of the BLR (\citealt{Chiang1996,Denney2010}). 

In practice, the measurement accuracy of SMBH mass depends on the factor $f_{\rm BLR}$ and the BLR velocity $V$ for a given H$\beta$ time lag. 
For the factor $f_{\rm BLR}$, which is calibrated from the known $M_{\bullet}-\sigma_{*}$ relation (e.g., \citealt{Woo2013}), where $\sigma_{*}$ is stellar velocity dispersion of bulge. 
In the case of the BLR is virialized and spherical, $f_{\rm BLR}\sim$1 for velocity characterized by FWHM (e.g., \citealt{Du2015}) 
and $f_{\rm BLR}\sim$5.5 for velocity characterized by the line dispersion $\sigma_{\rm line}$ (\citealt{Woo2013}). 
Based on the line dispersion from the rms spectrum, \cite{Grier2013b} gave an averaged value of $f_{\rm BLR}=4.31\pm1.05$ for RM sample. 
It should be noted that these values did not take the bulge type into consideration. 
\cite{Ho2014} evaluated the $f_{\rm BLR}$ by taking into account the bulge types of the host galaxies, 
and suggested that the $f_{\rm BLR}$ for the velocity characterized by the line dispersion approximates to 4$\sim$5~times larger than the $f_{\rm BLR}$ for the velocity characterized by FWHM. 
For the BLR velocity $V$, which can be characterized by either the FWHM or the line dispersion of the broad emission lines (such as H$\beta$), 
and both values can be measured from the mean and rms spectrum. 
From all the line-width measurements of H$\beta$ for Mrk~817 and NGC~7469 (Table~\ref{tab_sum}), 
we find that the FWHM(mean)$\simeq$(1.96$\pm$0.56)$\sigma_{\rm line}$(mean), and FWHM(rms)$\simeq$(1.93$\pm$0.68)$\sigma_{\rm line}$(rms), 
where ``(mean)" and ``(rms)" mean that the parameters are measured from the mean and rms spectrum, respectively. 
This line-width relation demonstrates that the FWHM-based virial product approximates to 2 times of the $\sigma_{\rm line}$-based virial product for a given H$\beta$ time lag. 

Because the bulge types of Mrk~817 and NGC~7469 are known, 
we prefer to adopt the value of $f_{\rm BLR}$ suggested by \cite{Ho2014} to estimate the SMBH mass for both targets. 
In Section~\ref{sec_bh}, we use $f_{\rm BLR}=3.2$ and the $\sigma_{\rm line}$ from the rms spectrum to estimate 
the SMBH masses and accretion rates for Mrk~817 and NGC~7469 (Table~\ref{tab_rm3}). 
To say the least, if we adopt an averaged value of $f=4.31$ given by \cite{Grier2013b} to yield a larger SMBH mass, which means that Mrk~817 accretes in a lower accretion rates, 
while NGC~7469 still accretes close to Eddington accretion rates 
($\dot{M}_{\bullet} \geq3 \times L_{\rm Edd}c^{-2}$) during \cite{Collier1998} and the present monitoring periods, according to the accretion criterion of \cite{Du2015}. 
As above discussed, if we use the FWHM and corresponding factor $f_{\rm BLR}$ (e.g., 0.7 suggested by \citealt{Ho2014}, or 1.0 adopted by \citealt{Du2015}) to estimate the SMBH mass, 
the SMBH masses of Mrk~817 and NGC~7469 are about 2 times smaller than the values listed in Table~\ref{tab_rm3}, 
and the accretion rates are about 4 times larger than the values listed in Table~\ref{tab_rm3}. 

\begin{figure}[ht!]
\centering
\includegraphics[angle=0,width=0.49\textwidth]{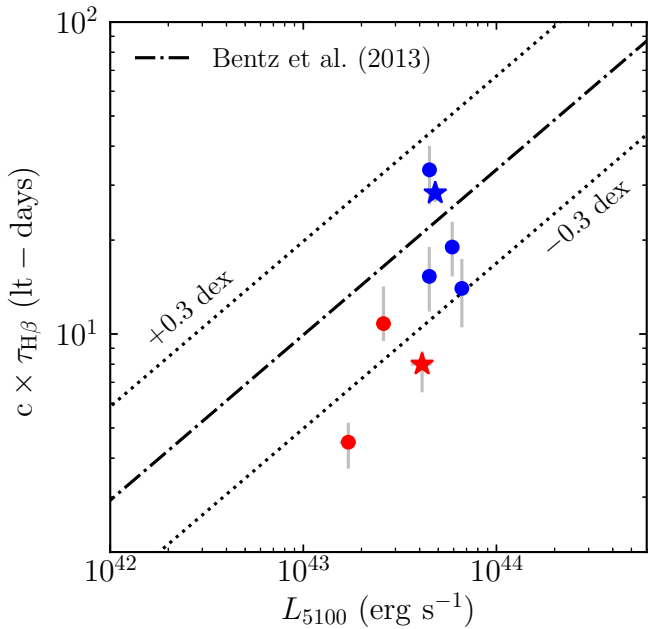}
\caption{\footnotesize
Optical continuum luminosity at 5100~\AA~and H$\beta$ time lag from all monitoring campaigns to date for Mrk~817 (in blue) and NGC~7469 (in red). 
The dot-dashed line is the Radius$-$Luminosity ($R_{\rm H\beta}-L_{\rm 5100}$) relation from \cite{Bentz2013}. 
The previous measurements for both targets show in symbol of circle, the measurements of this work show in ``$\bigstar$".} 
\label{fig_rl}
\end{figure}

\subsection{Scatter of $R_{\rm H\beta}-L_{\rm 5100}$ relation} 
We compare the H$\beta$ time lags and optical continuum luminosities at 5100~\AA~($L_{\rm 5100}$) from the present campaign with those from previous RM campaigns in Figure~\ref{fig_rl}, 
these parameters are listed in Table~\ref{tab_rm3}. 
The figure shows that all measurements of Mrk~817 follow \cite{Bentz2013}'s $R_{\rm H\beta}-L_{\rm 5100}$ relation within the scatter of 0.3~dex, 
and two out of three measurements of NGC~7469 from \cite{Collier1998} and the present campaign lie 0.3~dex below the \cite{Bentz2013}'s $R_{\rm H\beta}-L_{\rm 5100}$ relation. 

For Mrk~817, the maximum difference of BLR size is $\Delta R_{\rm BLR} \thickapprox 19$~days (see Table~\ref{tab_rm3}) in the five RM campaigns, 
while the maximum difference of optical luminosity is $\Delta {\rm log}~L_{\rm 5100}{\rm [erg~s^{-1}]} \thickapprox 0.17$, 
that is the optical flux of high-luminosity state is just $\sim$1.4~times of low-luminosity state. 
This is different with the multi-season RM measurements of NGC~5548 (\citealt{Lu2016}) and Mrk~79 (\citealt{Lu2019a}). 
In NGC~5548, the maximum difference of BLR size is $\Delta R_{\rm BLR}\thickapprox 22$~days, which is similar to Mrk~817, 
but its maximum difference of optical luminosity is $\Delta {\rm log}~L_{\rm 5100}{\rm [erg~s^{-1}]} \thickapprox 0.82$, indicating that 
the optical flux of NGC~5548 in high-luminosity state is more than $\sim$8~times larger than in low-luminosity state. 
What's more interesting is that, Mrk~817 and NGC~5548 have different properties in the X-ray and ultraviolet bands (e.g., \citealt{Winter2011,Morales2019}), 
that is the ultraviolet emissions are uncorrelated and correlated with X-ray for Mrk~817 and NGC~5548, respectively. 
These differences may suggest the ionized sources of BLR in different AGNs have different properties, 
which could act on the RM measurements, and then cause the scatter of $R_{\rm H\beta}-L_{\rm 5100}$ relation. 
For example, a transient ionised obscurer associated with an accretion disc wind can explain the BLR holiday in NGC 5548 \citep{Dehghanian2019a,Dehghanian2019b}. 

For NGC~7469, the discussion of Section~\ref{sec_dis2} demonstrates that it is a super-Eddington accretion AGN (at least approximates to Eddington accretion). 
According to the results of \cite{Du2015,Du2018}, there is a strong correlation between time lag shortening in SEAMBHs\footnote{
Super-Eddington Accreting Massive Black Holes, which is an AGN sample with high accretion rates from a large reverberation mapping campaign (see \citealt{Du2014}).} and accretion rates. 
\cite{Wang2014} put forward a possible explanation for time lag shortening. 
Specifically, slim accretion disc with super-Eddington accretion rates produce strong self-shadowing effects 
so that the BLR is thereby divided into a shadowed region and an unshadowed region. 
Since the shadowed region receives fewer ionizing photons, its size shrinks, leading to a shortened time lag. 

In addition, another possible explanation is that, as \cite{Fonseca2020} suggestion, 
the significant variation of the ionizing radiation, 
which may be related to the black hole spin and accretion rates, 
should cause the scatter of $R_{\rm H\beta}-L_{\rm 5100}$ relation, 
further developing the multi-season RM campaign for more famous AGNs could help us to understand the variations 
of BLR with the varying ionizing radiation (e.g., \citealt{Lu2016,Lu2019a}). 

\section{Summary}
\label{sec_con}
In this work, we present a new reverberation-mapping measurements of Mrk~817 and NGC~7469. 
Spectroscopic monitoring was fully undertaken with the Lijiang 2.4~m telescope, 
the median sampling interval is 2.0 days for Mrk~817 and 1.0 days for NGC~7469.  
Previous researches found that the ultraviolet emission of Mrk~817 is not correlation with X-ray, 
while the observation properties of X-ray and ultraviolet fully accord with the re-processing model of accretion disc for NGC~7469. 
Based on this campaign, we obtain the following results. 
\begin{enumerate}
\item 
The time lags of the broad emission lines including 
H$\beta$, H$\gamma$, He~{\sc ii} and He~{\sc i} for Mrk~817 and NGC~7469, and including Fe~{\sc ii} for Mrk~817 
with respect to the varying AGN continuum at 5100~\AA~are measured. 
These new measurements demonstrate that the BLR is the result of radial ionization stratification, which is consistent with previous findings. 
\item
We use the line dispersions and time lags  of all available broad emission lines to examine the virial relationship of Mrk~817 and NGC~7469, 
where the line dispersions are measured from the rms spectrum, and time lags are obtained from the centroid of CCF, 
and find that the BLR dynamics of Mrk~817 and NGC~7469 are consistent with the virial prediction. 
\item
Using the line dispersion from the rms spectrum and time lag of H$\beta$ and in combination with dimensionless virial factor {\bf ($f_{\rm BLR}=3.2$)} of pseudo-bulge, 
we measure SMBH mass of $M_{\bullet}=8.12^{+0.73}_{-0.67}\times10^{7}M_{\odot}$ for Mrk~817, 
and $M_{\bullet}=1.10^{+0.12}_{-0.21}\times10^{7}M_{\odot}$ for NGC~7469, 
which are consistent with the previous RM estimates within the uncertainties for each case. 
Especially, the SMBH mass of NGC~7469 obtained from the present campaign is consistent with 
estimator ($M_{\bullet}=1.78\times10^{7}M_{\odot}$) from the dynamical method of atomic and molecular emission lines. 
Combining the optical luminosity at 5100~\AA~and SMBH mass of the present campaign, 
we estimate accretion rates of $\dot{M}_{\bullet}=0.16~L_{\rm Edd}~c^{-2}$ for Mrk~817, 
and $\dot{M}_{\bullet}=6.72~L_{\rm Edd}~c^{-2}$ for NGC~7469. 
\item
We simultaneously recover the velocity-resolved lag profiles of the broad emission lines including 
H$\gamma$, H$\beta$, and He~{\sc i} for Mrk~817, and including H$\gamma$ and H$\beta$ for NGC~7469 for the first time, 
which show almost the same kinematic signatures that the time lags in the red wing are slightly larger than the time lags in the blue wing. 
This means that the BLR in both targets could has a part of outflowing components during the monitoring period. 
Although both targets ever observed outflow absorbers in the ultraviolet spectrum, 
the possible connections between these outflow absorbers and the kinematic signatures of BLR observed in the present campaign are still unclear. 
\item
We find that all H$\beta$ measurements of Mrk~817 follow the \cite{Bentz2013}'s $R_{\rm H\beta}-L_{\rm 5100}$ relation within the scatter of 0.3~dex, 
and two out of three H$\beta$ measurements of NGC~7469 from \cite{Collier1998} and the present campaign lie 0.3~dex below the \cite{Bentz2013}'s $R-L$ relation. 
We discuss the scatter of $R_{\rm H\beta}-L_{\rm 5100}$ relation. 
\end{enumerate}

\acknowledgements{
We thank the referee for useful reports that improved the manuscript. 
This work is supported by the National Natural Science Foundation of China (NSFC; 12073068, 11991051, 11703077, 11873048, 11803087, and U1931131). 
K.X.L. acknowledges financial support from the Yunnan Province Foundation (202001AT070069), 
and from the Light of West China Program provided by Chinese Academy of Sciences (Y7XB016001). 
L.X. acknowledges financial support from the Light of West China Program provided by Chinese Academy of Sciences (Y8XB018001). 
We acknowledge the support of the staff of the Lijiang 2.4 m telescope. 
Funding for the telescope has been provided by the CAS and the People’s Government of Yunnan Province. 
} 


\end{document}